\documentclass[12pt,longbibliography]{article}
\usepackage{amsmath,amssymb,amsthm,amsxtra,overpic,bbm,bm,epsfig,subfigure}
\usepackage{multirow} 
\usepackage{booktabs}
\usepackage{hyperref}
\usepackage{mathrsfs}
\usepackage{graphicx}
\usepackage{multirow}
\usepackage{makecell}
\usepackage{verbatim}
\usepackage{color}
\usepackage{comment}
\usepackage{epstopdf}
\usepackage{float}
\usepackage{amsmath}
\usepackage{amssymb}
\usepackage{tikz}
\usepackage{wrapfig}
\usepackage{tikz-feynman}
\usepackage{verbatim}
\numberwithin{equation}{section}
\usepackage{cite}
\usepackage{hyperref}
\usepackage{url}
\usepackage{slashed,stmaryrd}

\newcommand*{\dif}{\mathop{}\!\mathrm{d}}

\begin{document}
	\vspace{0.2cm}
\begin{center}
	{\Large\bf Measuring Solar neutrino Fluxes in Direct Detection Experiments in the Presence of Light Mediators}
\end{center}
\vspace{0.2cm}

\begin{center}
	{\bf Shuo-yu Xia}~$^{a,~b}$~\footnote{E-mail: xiashuoyu@ihep.ac.cn},
	\quad
	\\
	\vspace{0.2cm}
	{\small
		$^a$Institute of High Energy Physics, Chinese Academy of Sciences, Beijing 100049, China\\
		$^b$School of Physical Sciences, University of Chinese Academy of Sciences, Beijing 100049, China}
\end{center}

\begin{abstract}
The potential of the dark matter direct detection experiments to provide independent measurements on solar neutrino fluxes in the Standard Model and in the presence of the light mediators is studied in this work. We also present the sensitivity of direct detection experiments on light mediators with solar neutrinos. We find that the sensitivities on $^8$B and pp neutrino fluxes can reach $\pm10\%$ with improved backgrounds and systematic uncertainties and they can be further pushed to $\pm3\%$ with a increased exposure. The constraints on light mediators can reach $\mathcal{O}(10^{-6})$ for the masses of scalar and vector mediators below 10 MeV. However, the presence of scalar or vector mediators could lead to shifts in the best fit value of $^8$B fluxes, which will increase the challenges in the precise measurements of solar neutrino fluxes with direct detection experiments.
\end{abstract}

\newpage
\section{Introduction}
Since solar neutrinos were first observed at the Homestake experiment~\cite{Davis:1968cp}, the detection of solar neutrinos has been achieved with various technologies including radio-chemistry~\cite{Cleveland:1998nv,PhysRevC.80.015807}, water Cherenkov~\cite{SNO:2002tuh,Super-Kamiokande:2001ljr}, liquid scintillator~\cite{Borexino:2017uhp} and direct detection~\cite{XENON:2024ijk,PandaX:2024muv}. These measurements of solar neutrinos provide numerous physics opportunities for research in fields such as neutrino physics, dark matter search and astrophysics. Solar neutrinos, produced in the core of the Sun via nuclear fusion, exit the Sun {with flux strengths} practically unaffected, making them valuable probes of the interior structure of the sun~\cite{PhysRevLett.12.300}. Additionally, the fluxes of solar neutrinos are closely tied to the fusion rate, offering insights into the Sun's chemical composition and potentially resolving the discrepancies between metallicities and helioseismic observation data~\cite{Bahcall:2004yr}. Meanwhile, being one of the most intensive natural neutrino sources at the Earth, solar neutrinos also serve as a probe for the physics beyond the Standard Model{~\cite{Li:2022jfl,Demirci:2023tui,Bolanos:2008km,Borexino:2017fbd,PandaX-II:2020udv,Ge:2020jfn,Long:2013ota,Gondolo:2008dd,Super-Kamiokande:2004wqk,Amaral:2020tga,Chen:2021uuw,Chen:2022zts,Liao:2017awz}}. Furthermore, the neutrino induced signals in direct detection experiments overlap with weakly interacting massive particle (WIMPs), making them important components of the neutrino floor~\cite{Boehm:2018sux,Gonzalez-Garcia:2018dep,Papoulias:2018uzy}. The impact of the neutrino floor could be mitigated with accumulation of precise measurements of solar neutrino fluxes~\cite{OHare:2021utq}.

Recently, XENONnT~\cite{XENON:2024ijk} and PandaX-4T~\cite{PandaX:2024muv} announced their latest results on the measurements of the solar $^8$B neutrino flux via coherent elastic neutrino nucleus scattering (CE$\nu$NS), highlighting the potential of direct detection experiments to be effective platforms for solar neutrino flux measurements. With dark matter direct detection experiments now entering the multi-ton scale, various experiments and proposals like PandaX-4T~\cite{PandaX-4T:2021bab,PandaX:2024muv}, XENONnT~\cite{XENON:2018voc,XENON:2024ijk}, DARWIN~\cite{DARWIN:2016hyl} and Darkside-20k~\cite{DarkSide-20k:2017zyg} can provide new opportunities on the neutrino detection. These experiments, designed to detect rare and low-energy dark matter signals, are typically located underground, well-shielded, and feature low detection thresholds with large target masses. These characteristics also make them suitable for the neutrino detection. Given the stable and abundant flux of solar neutrinos, they will induce a considerable number of events at low energy and the coherent effect in CE$\nu$NS further increases event rates, making detecting solar neutrinos via nuclear recoils a promising physics opportunity. With ongoing advancements in direct detection technologies~\cite{Billard:2021uyg,Liu:2017drf,Schumann:2019eaa,Cooley:2022ufh,Klein:2022lrf}, increasingly precise measurements on solar neutrinos are anticipated in the coming decade.

The potential beyond Standard Model (BSM) sectors of neutrino physics draws much attention and considerable efforts have been focused on these fields including the neutrino electromagnetic properties~\cite{Cadeddu:2018dux,Cadeddu:2019eta,Kim:2021lun,Li:2022bqr}, nonstandard neutrino interactions~\cite{Coloma:2017ncl,Liao:2017uzy,Denton:2018xmq,AristizabalSierra:2018eqm,Giunti:2019xpr} and light mediators~\cite{Farzan:2018gtr,Abdullah:2018ykz,Ge:2017mcq,Cadeddu:2020nbr}. With intensive and stable fluxes, solar neutrinos serve as effective probes of BSM physics with a substantial number of low energy events in direct detection experiments~\cite{Cerdeno:2016sfi,Majumdar:2021vdw,Li:2022jfl}. However, the BSM physics may lead to excess or deficiency in solar neutrino fluxes once the corresponding contribution to the event rates cannot be distinguished with those of the standard ones with statistical methods since only the recoil signals are observable in such scenario. In this work, we consider the measurements of solar neutrino fluxes in the presence of the light mediators of universal scalar, vector and axial vector types on a series of simplified experimental setups. We demonstrate the potential of xenon-based direct detection experiments to independently measure solar neutrino fluxes in the Standard Model and the impact of light mediators on these measurements. We also present the independent constraints on light mediator models based on solar neutrino observations on our simplified experimental setups.

The plan of this work is as follows. In Section 2, we provides an overview of the theoretical framework for neutrino scatterings with nucleus and electrons in the Standard Model and in the presence of different light mediators. In Section 3, we illustrate the setups of the experimental scenarios and statistical analysis methods. In Section 4, we show the numerical results of our analysis and corresponding discussions. In Section 5, we give the concluding remarks.

\section{Theoretical Framework}
In this section, we will illustrate the theoretical framework of neutrino scatterings with nuclei and electron in the Standard Model and in the presence of different light mediators. We will also briefly comment on the signals induced by solar neutrino fluxes in direct detection experiments. 

\subsection{Coherent Elastic Neutrino Nucleus Scattering}
The cross section for CE$\nu$NS between a neutrino $\nu_\alpha$ with flavor $\alpha$ and a nucleus $\mathcal{N}$ in the SM can be written as~\cite{Drukier:1984vhf,Barranco:2005yy,Patton:2012jr,Li:2022jfl} 
\begin{equation}
\begin{aligned}
\frac{\dif{\sigma^{\nu_\alpha-\mathcal{N}}_{\mathrm{SM}}}}{\dif{T}_{\rm{NR}}}(E_{\nu},T_{\rm{NR}})=&\frac{G_{F}^2  M}{\pi}\left[\left(1-\frac{MT_{\rm{NR}}}{2E_{\nu}^2}+\frac{T_{\rm{NR}}}{E_{\nu}}\right)(Q^{V}_{\mathrm{SM}})^2\right.\\
&\left.+\left(1+\frac{MT_{\rm{NR}}}{2E_{\nu}^2}+\frac{T_{\rm{NR}}}{E_{\nu}}\right)(Q^{A}_{\mathrm{SM}})^2-2\left(\frac{T_{\rm{NR}}}{E_{\nu}}\right)Q^{V}_{\mathrm{SM}}Q^{A}_{\mathrm{SM}}\right]+\mathcal{O}\left({\frac{T_{\rm{NR}}^2}{E^2_{\nu}}}\right)\,,
\end{aligned}
\label{eq:cs_cevns_sm} 
\end{equation}
where $T_{\rm{NR}}$ is the kinetic energy of nuclear recoil and $E_\nu$ is the incident neutrino energy. The proton number, neutron number and mass of the nucleus $\mathcal{N}$ are denoted as $Z$, $N$ and $M$ respectively. The vector and axial vector weak charge $Q^{V}_{\mathrm{SM}}$ and $Q^{A}_{\mathrm{SM}}$ are given in the SM as~\cite{Li:2022jfl}
\begin{equation}
Q^{V}_{\mathrm{SM}}=
[g_{V}^{p} Z +g_{V}^{n} N]F_{V}\left(|\vec{q}|^{2}\right)\quad{\rm and}\quad
Q^{A}_{\mathrm{SM}}=
[g_{A}^{p}(Z^{+}-Z^{-}) +g_{A}^{n}(N^{+}-N^{-})]F_{A}\left(|\vec{q}|^{2}\right),
\label{eq:charge_sm} 
\end{equation}
where $|\vec{q}|^{2}=2MT_{\rm{NR}}$ and $F_{V}\left(|\vec{q}|^{2}\right)$ and $F_{A}\left(|\vec{q}|^{2}\right)$  are the vector and axial vector form factors of nucleus $\mathcal{N}$. In this work, we consider liquid xenon as the target material and the proton rms radii $R_Z=4.7808$ fm~\cite{Angeli:2013epw} and the neutron rms radii $R_N=5.08$ fm~\cite{Hoferichter:2020osn} are adopted. $g_{V}^{p}$ and $g_{V}^{n}$ are the vector neutrino-proton and neutrino-neutron couplings in the SM and we employ the radiative correction results under the $\overline{\rm MS}$ scheme as~\cite{Erler:2013xha,PhysRevD.98.030001} 
\begin{equation}
\begin{aligned}
g_{V}^{p}(\nu_e)=0.0401\,,&\quad\quad g_{V}^{p}(\nu_\mu)=0.0318\,,\\
g_{V}^{p}(\nu_\tau)=0.0275\,,&\quad\quad g_{V}^{n}=-0.5094\,.\\
\end{aligned}
\end{equation}
The axial vector neutrino-proton and neutrino-neutron couplings $g_{A}^{p}$ and $g_{A}^{n}$ are detailed in Ref.~\cite{Cirelli:2013ufw} and $Z^{\pm}$ and $N^{\pm}$ are notations for the numbers of protons and neutrons with spin up (spin down) respectively. Therefore, the axial vector components in the CE$\nu$NS process is coherently coupled to the spin of the target nucleus. In this work, we choose xenon as the target material and only $^{131}$Xe and $^{129}$Xe have non-zero spins of 3/2 and 1/2 respectively. The contribution of the axial vector component in CE$\nu$NS process is sub-dominant as shown in Eq.~\ref{eq:cs_cevns_sm} and the abundances of $^{131}$Xe and $^{129}$Xe in natural xenon is about 21\% and 26\%, which makes the axial vector contribution in xenon direct detection experiments even smaller.

The neutrinos scatter with the nucleus coherently at low energy and the cross section is enhanced with the square of vector charge as shown in Eq.~\ref{eq:charge_sm}, among which the contribution from neutron number is dominant. This makes the CE$\nu$NS process an efficient tool to detect low energy neutrino scatterings and a significantly increased signal rate is expected compared to neutrino electron scatterings. However, the minimum energy for an incident neutrino to induce a CE$\nu$NS event with a recoil energy of $T_{\rm{NR}}$ is $E_{\rm{min}}\simeq \sqrt{MT_{\rm{NR}}/2}$ and this makes it very difficult to observe CE$\nu$NS events from solar neutrinos in direct detection experiments. Present xenon direct detection experiments like PandaX-4T~\cite{PandaX:2024muv} and XENONnT~\cite{XENON:2024ijk} present thresholds less than 1 keV$_{\rm{NR}}$ for nuclear recoils and it makes it possible to observe CE$\nu$NS events induced by $^8$B solar neutrinos. The $^8$B neutrinos make the dominant contribution to the solar neutrino nuclear recoil signals at several keV$_{\rm{NR}}$ on present direct detection experiments. Although the hep neutrinos can also induce nuclear recoil events at several keV$_{\rm{NR}}$, the corresponding event rate is too low to be observed in present ton-scale experiments. It may become feasible to observe hep neutrino events in future experiments with increased exposures and at about 4 to 5 keV$_{\rm{NR}}$ the hep nuclear recoil events can be measured more precisely without the $^8$B signals.

\subsection{Elastic Neutrino Electron Scattering}
The cross section for elastic neutrino electron scattering (E$\nu$ES) between 
a neutrino $\nu_\alpha$ with flavor $\alpha$ and an electron in the SM can be written as
\begin{equation}
\frac{\dif{\sigma^{\nu_\alpha-e}_{\mathrm{SM}}}}{\dif{T_e}}(E_{\nu},T_e)=Z\frac{G_{\rm{F}}^2m_e}{2\pi}\left[ (g_V^{\nu_\alpha}+g_A^{\nu_\alpha})^2+(g_V^{\nu_\alpha}-g_A^{\nu_\alpha})^2\left( 1-\frac{T_e}{E_\nu}\right)^2-\left( (g_V^{\nu_\alpha})^2-(g_A^{\nu_\alpha})^2\right)\frac{m_eT_e}{E^2_\nu}  \right] \,,
\end{equation}
where $m_e$ is the mass of the electron and $G_{\rm{F}}$ is the Fermi Constant. In this work, we adopt the Free Electron Approximation and $Z$ is the number of electrons in a target atom. For electron flavor neutrinos the E$\nu$ES can occur via either neutral or charged current interactions while $\mu$ and $\tau$ flavor neutrinos can scatter with electrons only through neutral current interactions. Therefore, the vector and axial vector neutrino-electron couplings differ for neutrinos of different flavors and are given by
\begin{equation}
	\begin{aligned}
		&g^{\nu_e}_V=2\sin^2\theta_W+1/2,\quad\quad g^{\nu_e}_A=1/2,\\
		&g^{\nu_{\mu,\tau}}_V=2\sin^2\theta_W-1/2,\quad\quad g^{\nu_{\mu,\tau}}_A=-1/2\,,\\ 
	\end{aligned}
\end{equation}
where the $\theta_W$ is the weak mixing angle. We adopt the value $\sin^2\theta_W=0.23857$ at zero momentum transfer in the $\overline{\rm MS}$ scheme. The minimum energy for an incident neutrino to induce a E$\nu$ES event with a recoil energy of $T_{e}$ is $E_{\rm{min}}= (T_e+\sqrt{T_e^2+2m_eT_e})/2$. This enables the observation of all solar neutrino fluxes in current direct detection experiments with a region of interest (ROI) of $\mathcal{O}(1)$ keV to $\mathcal{O}(100)$ keV. However, the E$\nu$ES cross section is much smaller than that of CE$\nu$NS due to the absence of coherent enhancement and the event rate is not very sensitive to the recoil energy $T_e$ at ROI of current direct detection experiments. The pp neutrino flux induces the dominant number of events at such energy range among solar neutrino fluxes and the uncertainties of the pp neutrino will significantly hinder the measurements of the other fluxes. In other electron scattering experiments like Borexino, the upper limit of ROI can extend to more than 1 MeV and the disappearance of pp signals will make the measurement of $^7$B, pep and CNO fluxes much easier. However, in liquid noble gas direct detection experiments, the background in this energy range is very high. The background discrimination at higher energies, such as Compton photons, is necessary for the extension of ROI in direct detection experiments in the future.

\subsection{Contribution Form Light Mediators}
In this work, we consider a scenario where different flavors of neutrinos and quarks universally couple to light mediators through general scalar, vector and axial vector interactions, which could be a good tool to explore the physics beyond Standard Model (BSM)in a model-independent manner~\cite{PhysRevD.98.075018,Lindner:2016wff}. We employ the effective field theory framework from ref.~\cite{Cirelli:2013ufw} to describe the low energy neutrino scattering in the presence of BSM couplings and light mediators and the extended Lagrangian for the flavor-universal scalar (S), vector (V) and axial vector (A) interactions can be written as{~\cite{Cirelli:2013ufw,AristizabalSierra:2018eqm,Lindner:2016wff}} 
{
\begin{equation}
\begin{aligned}
\mathcal{L}_{\text {S}}&=\phi\left(g_{\phi}^{f S} \bar{f} f+g_{\phi}^{\nu S} \bar{\nu}_{R} \nu_{L}+\rm{h.c.}\right),\\
\mathcal{L}_{\text {V}}&=Z_{\mu}^{\prime}\left(g_{Z^{\prime}}^{f V} \bar{f} \gamma^{\mu} f+g_{Z^{\prime}}^{\nu V} \bar{\nu}_{L} \gamma^{\mu} \nu_{L}\right), \\
\mathcal{L}_{\text {A}}&=Z_{\mu}^{\prime}\left(g_{Z^{\prime}}^{f A} \bar{f} \gamma^{\mu} \gamma^{5}f+g_{Z^{\prime}}^{\nu A} \bar{\nu}_{L} \gamma^{\mu} \gamma^{5} \nu_{L}\right),
\end{aligned}
\label{lagragian:nsi}
\end{equation}}
where $g_\phi^{qS}$ and $g_\phi^{\nu S}$ are the scalar couplings between light mediator $\phi$ and fermions $f=(u,d,e)$ and neutrinos, while $g_{Z^{\prime}}^{q V(A)}$ and $g_{Z^{\prime}}^{\nu V(A)}$ are the vector (axial vector) couplings between light mediator $Z^{\prime}$ and fermions $f=(u,d,e)$ and neutrinos. $\phi$ and $Z^{\prime}$ are the corresponding light mediators for scalar and vector (axial vector) interactions respectively. It should be noted that couplings in the three interactions derived from Lagrangian in Eq.~\ref{lagragian:nsi} can only contribute to the cross section in pairs since the light mediators couple to both neutrinos and quarks/electrons in scattering processes. Therefore, we write the corresponding couplings as effective couplings as $(g_{eff}^S)^2=g_\phi^{qS}g_\phi^{\nu S}$ and {$(g_{eff}^{V(A)})^2=g_{Z^{\prime}}^{q V(A)}g_{Z^{\prime}}^{\nu V(A)}$} for scalar and vector (axial vector) interactions. Although the light mediator models will modify the matter effect on neutrinos as they travel through the Sun, the oscillation behaviour of the solar neutrino will not change since in this work all the couplings are universal for neutrinos of different flavors and their corresponding contributions can be eliminated by a phase shift similar to the neutral current contributions from neutrino-neutron coherent forward scattering.

Since the Standard Model neutrino scattering does not include a scalar weak charge, the contribution from scalar interactions is incoherent with the standard cross section. The cross section for CE$\nu$NS in the presence of scalar interaction can be written as
\begin{equation}
\dfrac{d\sigma^{\nu_\alpha-\mathcal{N}}_{\mathrm{SM+S}}}{d T_{\rm{NR}}}(E_{\nu},T_{\rm{NR}})
=\frac{\dif{\sigma^{\nu_\alpha-\mathcal{N}}_{\mathrm{SM}}}}{\dif{T}_{\rm{NR}}}(E_{\nu},T_{\rm{NR}})+\frac{M^{2}}{4 \pi}
\frac{T_{\rm{NR}}}{E_{\nu}^{2}}
\frac{({Q}^{S}_{\phi})^{2}}{\left(|\vec{q}|^{2}+M_{\phi}^{2}\right)^{2}}\,,
\label{cs_scalar}
\end{equation}
where $M_{\phi}$ is the mass of the scalar light mediator and the scalar weak charge is given as~\cite{AristizabalSierra:2019ykk,Majumdar:2021vdw,Li:2022jfl}
\begin{equation}\label{Q_scalar}
{Q}^{S}_{\phi}=\left[Z\sum_{q=u, d, s}  \frac{m_{p}}{m_{q}} f_{T_{q}}^{p}+N  \sum_{q=u, d, s} \frac{m_{n}}{m_{q}} f_{T_{q}}^{n}\right](g_{eff}^S)^2,
\end{equation}
where $m_q$, $m_p$ and $m_n$ are the mass of quarks, proton and neutron. The hadronic form factors $f_{T_{q}}^{p,n}$ are obtained from the chiral perturbation theory~\cite{AristizabalSierra:2019zmy,Ellis:2018dmb,Hoferichter:2015dsa}. The cross section for E$\nu$ES in the presence of scalar interaction can be written as~\cite{Link:2019pbm}
\begin{equation}
\frac{\dif{\sigma^{\nu_\alpha-e}_{\mathrm{SM+S}}}}{\dif{T_e}}(E_{\nu},T_e)= \frac{\dif{\sigma^{\nu_\alpha-e}_{\mathrm{SM}}}}{\dif{T_e}}(E_{\nu},T_e)+\left[\frac{(g_{eff}^S)^4}{4\pi (2m_eT_e+M_{\phi}^2)^2} \right]\frac{m_e^2 T_e}{E^2_\nu} \,,
\end{equation}

For the light mediators with vector or axial vector interactions, the corresponding contributions are coherent with the vector or axial vector components in the Standard Model neutrino scattering. For the CE$\nu$NS process, the vector (axial vector) interactions modify the vector (axial vector) weak charge in Eq.~\ref{eq:cs_cevns_sm} with additional terms~\cite{Li:2022jfl}
\begin{equation}\label{Q_VA}
\begin{aligned}
&Q^{V}_{\mathrm{SM}}\rightarrow Q^{V}_{\mathrm{SM+V}}=
Q^{V}_{\mathrm{SM}} - 
\frac{3(g_{eff}^V)^2(Z + N)}
{\sqrt{2} G_{F}\left(|\vec{q}|^{2}+M_{Z^{\prime}}^{2}\right)} 
\,,\\
Q^{A}_{\mathrm{SM}}\rightarrow Q^{A}_{\mathrm{SM+A}}=&
Q^{A}_{\mathrm{SM}} + 
\frac{(g_{eff}^A)^2\left[(\sum_{q}\Delta^{(p)}_{q})(Z^{+}-Z^{-}) + (\sum_{q}\Delta^{(n)}_{q}) (N^{+}-N^{-})\right]}
{\sqrt{2} G_{F}\left(|\vec{q}|^{2}+M_{Z^{\prime}}^{2}\right)} 
\,,
\end{aligned}
\end{equation}
where $M_{Z^{\prime}}$ is the mass of the vector (axial vector) light mediator. Similar to the axial vector weak charge in the Standard Model, the non-standard axial vector weak charge also depends on the spin of the nucleus with $\sum_{q}\Delta^{(p)}_{q}=\sum_{q}\Delta^{(n)}_{q}\simeq0.3$~\cite{Cerdeno:2016sfi}. For the E$\nu$ES process, the cross section in the presence of the vector (axial vector) interaction can be obtained with modifications on the vector (axial vector) neutrino-electron couplings
\begin{equation}
g_{V(A)}^{\nu_\alpha}\rightarrow {g_{V(A)}^{\nu_\alpha}}^{\prime}=g_{V(A)}^{\nu_\alpha}+\frac{(g_{eff}^{V(A)})^2}{4\sqrt{2}G_{\rm{F}}(2m_eT_e+m^2_{Z^{\prime}})}\,,
\end{equation} 
It should be noted that the contributions from vector and axial vector interaction E$\nu$ES are in the same form and it can be difficult to distinguish between them with only neutrino-electron scattering. However, for CE$\nu$NS the cross section is much more sensitive to the vector coupling than the axial vector coupling since the vector contribution coherently couples to the nucleon number, which is of the order of $\mathcal{O}(10)$ - $\mathcal{O}(100)$, while the axial-vector contribution couples to the spin number, which is of the order of $\mathcal{O}(1)$.

\section{Statistical Framework}
In this section, we will present the statistical method and setups of the simplified experimental scenarios we employ to analyze the sensitivity of future direct detection experiments on the solar neutrino fluxes and the light mediators.

\subsection{Statistical Method}
In this work we discuss the CE$\nu$NS and E$\nu$ES signals induced by solar neutrinos in liquid xenon direct detection experiments in the Standard Model and in the presence of different light mediators. The predicted CE$\nu$NS event number in each bin can be written as
\begin{equation}
N^{\nu-\mathcal{N}}_{i} = \frac{\epsilon}{M} \int_{(T_{\rm{NR}}^{\rm min})_i}^{(T_{\rm{NR}}^{\rm max})_i}\eta^{\nu-\mathcal{N}}(T_{\rm{NR}}) \dif{T_{\rm{NR}}}\sum_j\sum_\alpha \int_{E_{{\rm min}}}^{E_{{\rm max}}} \dif{E_{\nu}} \cdot P_{e\alpha}\Phi_j (E_{\nu}) \frac{\dif{\sigma^{\nu_\alpha-\mathcal{N}}}}{\dif{T}_{\rm{NR}}}(E_{\nu},T_{\rm{NR}})\,,
\end{equation}
where $\epsilon$ is the exposure of the experiment and $M$ is the mass of the target nucleus. In this work, we apply a weighted average on the event rates based on the natural abundance of different isotopes of xenon. $(T_{\rm{NR}}^{\rm max})_i$ and $(T_{\rm{NR}}^{\rm min})_i$ are the upper and lower limits of each bin in keV$_{\rm NR}$ and $\eta^{\nu-\mathcal{N}}(T_{\rm{NR}})$ is the total detection efficiency of the nuclear recoil signals. We adopt the total detection efficiency from XENONnT~\cite{XENON:2024ijk} and the threshold for our analysis on CE$\nu$NS is 0.5 keV$_{\rm{NR}}$. $\Phi_j (E_{\nu})$ with $j=(^7{\rm Be}\,,^8{\rm B}\,,pp\,,pep\,,hep\,,{\rm CNO})$ are the reference solar neutrino fluxes from the standard solar model B23~\cite{Gonzalez-Garcia:2023kva} shown in Tab.~\ref{table:flux} and $P_{e\alpha}$ is the oscillation factor for neutrinos of flavor $\alpha$. $E_{\rm{min}}\simeq \sqrt{MT_{\rm{NR}}/2}$ is the minimum energy for neutrinos to induce nuclear recoil signals and $E_{\rm{max}}$ is the maximum energy for each flux. Only $^8$B and $hep$ neutrinos contribute to the CE$\nu$NS events in our ROI due to the minimum energy limits. Similarly, the predicted E$\nu$ES event number in each bin can be written as
\begin{equation}
N^{\nu-e}_{i} = \frac{\epsilon}{M} \int_{(T_{e}^{\rm min})_i}^{(T_{e}^{\rm max})_i}\eta^{\nu-e}(T_{e}) \dif{T_{e}}\sum_j\sum_\alpha  \int_{E_{{\rm min}}}^{E_{{\rm max}}} \dif{E_{\nu}} \cdot  P_{e\alpha}\Phi_j (E_{\nu})\frac{\dif{\sigma^{\nu_\alpha-e}}}{\dif{T}_{e}}(E_{\nu},T_{e})\,,
\end{equation}
where we set the ROI as 1-140 keV to avoid large background rates at higher energy and we adopt the electron recoil detection efficiency in XENONnT~\cite{XENON:2022ltv} as the efficiency $\eta^{\nu-e}(T_{e})$ for analysis. The minimum neutrino energy to induce electron recoil is $E_{\rm{min}}= (T_e+\sqrt{T_e^2+2m_eT_e})/2$ and this makes all solar neutrino components are capable of inducing electron recoil in our ROI.

Present liquid xenon dual-phase time projection chambers can already distinguish nuclear recoils and electron recoils very efficiently with different ratios of scintillation and ionization yields~\cite{XENON:2020kmp}. Therefore, in this work we analyze the nuclear and electron recoil signals independent of each other. We employ the least squares method with the Poissonian likelihood function and the total $\chi^2$ function is the sum of the $\chi^2$ functions from analysis of CE$\nu$NS and E$\nu$ES functions respectively
\begin{equation}
    \chi^2=\chi^2_{\rm CE\nu NS}+\chi^2_{\rm E\nu ES}\,.
\end{equation}
In this work, we will consider three kind of independent measurements on the solar neutrino fluxes with certain direct detection experiment setups:
\begin{itemize}
    \item First, we will consider the potential of xenon-based direct detection experiments to independently constrain the neutrino flux strengths. In this scenario, the $\chi^2$ functions for the analysis of CE$\nu$NS and E$\nu$ES events can be written as
    \begin{equation}
        \begin{aligned}
        \chi^2_{\rm CE\nu NS}&=2\sum_{i=1}^{n}\Bigg[(1+\epsilon^{\nu-\mathcal{N}}_{\rm exp})(N^{\nu-\mathcal{N}}_{i})^{{\rm pred}}(\lbrace(1+\epsilon_j)\Phi_j\rbrace )-(N^{\nu-\mathcal{N}}_{i})^{{\rm exp}}\\
        &+(N^{\nu-\mathcal{N}}_{i})^{{\rm exp}}\rm{ln}\left(\frac{(N^{\nu-\mathcal{N}}_{i})^{{\rm exp}}}{(1+\epsilon^{\nu-\mathcal{N}}_{\rm exp})(N^{\nu-\mathcal{N}}_{i})^{{\rm pred}}(\lbrace(1+\epsilon_j)\Phi_j\rbrace )}\right)\Bigg]+\left(\frac{\epsilon^{\nu-\mathcal{N}}_{\rm exp}}{\sigma_{\rm exp}} \right)^2 \,,\\
        \chi^2_{\rm E\nu ES}&=2\sum_{i=1}^{n}\Bigg[(1+\epsilon^{\nu-e}_{\rm exp})(N^{\nu-e}_{i})^{{\rm pred}}(\lbrace(1+\epsilon_j)\Phi_j\rbrace )-(N^{\nu-e}_{i})^{{\rm exp}}\\
        &+(N^{\nu-e}_{i})^{{\rm exp}}\rm{ln}\left(\frac{(N^{\nu-e}_{i})^{{\rm exp}}}{(1+\epsilon^{\nu-e}_{\rm exp})(N^{\nu-e}_{i})^{{\rm pred}}(\lbrace(1+\epsilon_j)\Phi_j\rbrace )}\right)\Bigg]+\left(\frac{\epsilon^{\nu-e}_{\rm exp}}{\sigma_{\rm exp}} \right)^2 \,,\\
        \end{aligned}
        \label{chi2:base}
    \end{equation}
    where $(N^{\nu-\mathcal{N}}_{i})^{{\rm exp(pred)}}$ and $(N^{\nu-e}_{i})^{{\rm exp(pred)}}$ represent the experimental (prediction) event numbers of CE$\nu$NS and E$\nu$ES based on the considered experimental setups and $(N^{\nu-\mathcal{N}}_{i})^{{\rm bkg}}$ and $(N^{\nu-e}_{i})^{{\rm bkg}}$ are the corresponding background event numbers. In this scenario all the events are induced by standard scattering processes. The experimental event numbers are calculated using the reference flux values presented in Tab.~\ref{table:flux}, while the prediction event numbers are based on modified fluxes of $(1+\epsilon_j)\Phi_j$. $\epsilon^{\nu-\mathcal{N}}_{\rm exp}$ and $\epsilon^{\nu-e}_{\rm exp}$ are the simplified nuisance parameters that quantify the total systematic uncertainties in the analysis of CE$\nu$NS and E$\nu$ES signals and $\sigma_{\rm exp}$ is the systematic uncertainty depend on the specific experimental setups. In this scenario we impose no constraint on solar neutrino fluxes to ensure that the results are independently achieved by the direct detection experiments we design.

    \begin{table}
    	\centering
    	\begin{tabular}{cc}
    		\toprule
    		Flux& $(\Phi_0)_j\,[{\rm cm}^{-2}{\rm s}^{-1}]$  \\
    		\midrule 
    		{$^7$Be}&4.93$\times10^{9}$\\
    		{$^8$B}&5.20$\times10^{6}$\\
    		$pp$&5.941$\times10^{10}$\\
    		$pep$&1.421$\times10^{8}$\\
    		$hep$&3.0$\times10^{4}$ \\
    		{$^{13}$N}&3.48$\times10^{8}$\\
    		{$^{15}$O}&2.53$\times10^{8}$\\
    		{$^{17}$F}&5.51$\times10^{7}$ \\
    		\bottomrule
    	\end{tabular}
    	\caption{The reference solar neutrino fluxes adopted from the Standard Solar Model B23~\cite{Gonzalez-Garcia:2023kva} for normalization in this work.}
    	\label{table:flux}
    \end{table}

    \item With the constraints obtained in the first scenario, we project the sensitivities of the light mediator models on the direct detection experiment setups using solar neutrino fluxes. The corresponding $\chi^2$ function for analyzing the CE$\nu$NS and E$\nu$ES signals in this scenario is expressed as Eq.~\ref{chi2:base} with additional pull terms introduced to incorporate the constraints obtained in the first scenario
    \begin{equation}
        \begin{aligned}
        \chi^2_{\rm CE\nu NS}&=2\sum_{i=1}^{n}\Bigg[(1+\epsilon^{\nu-\mathcal{N}}_{\rm exp})(N^{\nu-\mathcal{N}}_{i})^{{\rm pred}}(\lbrace(1+\epsilon_j)\Phi_j\rbrace )-(N^{\nu-\mathcal{N}}_{i})^{{\rm exp}}\\
        &+(N^{\nu-\mathcal{N}}_{i})^{{\rm exp}}\rm{ln}\left(\frac{(N^{\nu-\mathcal{N}}_{i})^{{\rm exp}}}{(1+\epsilon^{\nu-\mathcal{N}}_{\rm exp})(N^{\nu-\mathcal{N}}_{i})^{{\rm pred}}(\lbrace(1+\epsilon_j)\Phi_j\rbrace )}\right)\Bigg]+\left(\frac{\epsilon^{\nu-\mathcal{N}}_{\rm exp}}{\sigma_{\rm exp}} \right)^2 +\sum_j \left(\frac{\epsilon_j}{\sigma_j} \right)^2\,,\\
        \chi^2_{\rm E\nu ES}&=2\sum_{i=1}^{n}\Bigg[(1+\epsilon^{\nu-e}_{\rm exp})(N^{\nu-e}_{i})^{{\rm pred}}(\lbrace(1+\epsilon_j)\Phi_j\rbrace )-(N^{\nu-e}_{i})^{{\rm exp}}\\
        &+(N^{\nu-e}_{i})^{{\rm exp}}\rm{ln}\left(\frac{(N^{\nu-e}_{i})^{{\rm exp}}}{(1+\epsilon^{\nu-e}_{\rm exp})(N^{\nu-e}_{i})^{{\rm pred}}(\lbrace(1+\epsilon_j)\Phi_j\rbrace )}\right)\Bigg]+\left(\frac{\epsilon^{\nu-e}_{\rm exp}}{\sigma_{\rm exp}} \right)^2 +\sum_j \left(\frac{\epsilon_j}{\sigma_j} \right)^2\,,\\
        \end{aligned}
        \label{chi2:nsi}
    \end{equation}
    where $\sigma_j$ is the corresponding uncertainties associated with the flux $\Phi_j$.
    
    \item In the presence of the light mediators, the BSM contribution to the solar neutrino event rate can also lead to excess or deficiency in solar neutrino flux measurements. Such excess (deficiency) cannot be resolved once the BSM contributions cannot be distinguished with those from the standard scatterings based on statistical methods like analysis of event spectrums. Therefore, in the third scenario we project the potential excess (deficiency) in the measurements of solar neutrino fluxes induced by the presence of the light mediators. The corresponding $\chi^2$ function for the analysis of CE$\nu$NS and E$\nu$ES signals can be written as
    \begin{equation}
        \begin{aligned}
        \chi^2_{\rm CE\nu NS}&=2\sum_{i=1}^{n}\Bigg[(1+\epsilon^{\nu-\mathcal{N}}_{\rm exp})(N^{\nu-\mathcal{N}}_{i})^{{\rm pred}}(\lbrace(1+\epsilon_j)\Phi_j\rbrace )-(N^{\nu-\mathcal{N}}_{i})^{{\rm exp}}_{\rm SM+X}\\
        &+(N^{\nu-\mathcal{N}}_{i})^{{\rm exp}}_{\rm SM+X}\rm{ln}\left(\frac{(N^{\nu-\mathcal{N}}_{i})^{{\rm exp}}_{\rm SM+X}}{(1+\epsilon^{\nu-\mathcal{N}}_{\rm exp})(N^{\nu-\mathcal{N}}_{i})^{{\rm pred}}(\lbrace(1+\epsilon_j)\Phi_j\rbrace )}\right)\Bigg]+\left(\frac{\epsilon^{\nu-\mathcal{N}}_{\rm exp}}{\sigma_{\rm exp}} \right)^2 \,,\\
        \chi^2_{\rm E\nu ES}&=2\sum_{i=1}^{n}\Bigg[(1+\epsilon^{\nu-e}_{\rm exp})(N^{\nu-e}_{i})^{{\rm pred}}(\lbrace(1+\epsilon_j)\Phi_j\rbrace )-(N^{\nu-e}_{i})^{{\rm exp}}_{\rm SM+X}\\
        &+(N^{\nu-e}_{i})^{{\rm exp}}_{\rm SM+X}\rm{ln}\left(\frac{(N^{\nu-e}_{i})^{{\rm exp}}_{\rm SM+X}}{(1+\epsilon^{\nu-e}_{\rm exp})(N^{\nu-e}_{i})^{{\rm pred}}(\lbrace(1+\epsilon_j)\Phi_j\rbrace )}\right)\Bigg]+\left(\frac{\epsilon^{\nu-e}_{\rm exp}}{\sigma_{\rm exp}} \right)^2 \,.\\
        \end{aligned}
        \label{chi2:shift}
    \end{equation}
    We utilize similar $\chi^2$ functions with Eq.~\ref{chi2:base} to show the potential excess (deficiency) induced by light mediators. To address this we include additional light mediator contributions into the experimental event numbers $(N^{\nu-\mathcal{N}}_{i})^{{\rm exp}}_{\rm SM+X}$ and $(N^{\nu-e}_{i})^{{\rm exp}}_{\rm SM+X}$ with ${\rm X=V\,,S\,,A}$ for vector, scalar and axial vector mediator models. To project the potential excess (deficiency) induced by light mediators we will compare the constraints and best-fit flux values achieved from Eq.~\ref{chi2:base} and Eq.~\ref{chi2:shift}.
\end{itemize}

\subsection{Experimental Scenarios}
Today the direct detection experiments have entered the phase of multi-ton scale and present experiments are already capable of detecting solar neutrinos. Recently, the $^8$B neutrino flux is measured with CE$\nu$NS in XENONnT~\cite{XENON:2024ijk} and PandaX-4T~\cite{PandaX:2024muv}. This shows the potential of xenon direct detection experiments to provide precise measurements of the solar neutrino fluxes. LUX-ZEPLIN~\cite{LZ:2022lsv} has also deployed a exposure of 60 live days with a fiducial mass of 5.5 t in 2022. The future flagship experiment DARWIN is expected to deploy an unprecedented target mass of 50 tons of liquid xenon, which could also be a good platform for solar neutrino measurements.

In today's direct detection experiments liquid xenon is a widely-used target material and xenon-based experiments have had great success for exploring low-energy physics~\cite{LZ:2015kxe,XENON:2024ijk,PandaX:2024muv,XENON:2018voc,Maity:2024aji}. Liquid xenon has several significant advantages as the target material in the direct detection of neutrinos. The nucleon number of xenon is very large and provide a strong coherent effect in CE$\nu$NS, which induces the dominant solar neutrino signals at several keV$_{\rm NR}$. Also, the radioactive backgrounds in liquid xenon are very low and the high density of liquid xenon provides effective shielding of neutrons and $\gamma$ rays. Moreover, the background signals and the detection efficiency of xenon-based direct detection experiments are well researched in earlier works~\cite{XENON:2024ijk,PandaX:2024muv,XENON:2020kmp,XENON:2018voc,PandaX-II:2016vec}.

Motivated by the above investigation, we shall consider the experimental scenarios listed in Tab.~\ref{tab_set} using liquid xenon as the target material. We consider two exposure scenarios of 100 t$\cdot$yr and 1000 t$\cdot$yr for future experimental setups with fiducial masses of tens and hundreds of tons, similar to future experiments like DARWIN~\cite{DARWIN:2016hyl} and ARGO-scaled xenon-based experiments.

In this work, we consider three background models $B_0$, $B_1$ and $B_2$. Model $B_0$ includes the realistic background for nuclear recoil in ref.~\cite{XENON:2020kmp} and the electron recoil background model in ref.~\cite{XENON:2022ltv} and the corresponding accidental coincidence (AC) background signals. Since the quantity of AC signals depends on specific experimental setups and analysis methods, we consider flat AC backgrounds of 3 t$^{-1}$yr$^{-1}$keV$^{-1}$ for nuclear recoil and 0.005 t$^{-1}$yr$^{-1}$keV$^{-1}$ for electron recoil in $B_0$ based on data-driven best-fit results in ref.~\cite{XENON:2024ijk} and ref.~\cite{XENON:2022ltv} respectively. Models $B_1$ and $B_2$ are flat background models designed to illustrate the impact of the improvement of technologies on background discrimination in future. Model $B_1$ includes flat backgrounds of 0.5 t$^{-1}$yr$^{-1}$keV$^{-1}$ for nuclear recoil and 10 t$^{-1}$yr$^{-1}$keV$^{-1}$ for electron recoil, which represents a feasible improvement over $B_0$ by less than an order of magnitude. In $B_1$, the nuclear recoil background rate is similar to the corresponding CE$\nu$NS event rate when considering the detector efficiency mentioned above and the electron recoil background rate is similar to the background rate of Xenon-nT~\cite{XENON:2022ltv} at a few tens of keV, but significant smaller at higher energy. Model $B_2$ includes flat backgrounds of 0.1 t$^{-1}$yr$^{-1}$keV$^{-1}$ for nuclear recoil and 1 t$^{-1}$yr$^{-1}$keV$^{-1}$ for electron recoil, which represents a significant improvement from $B_0$ that can be realized in decades.

We consider 8 experiment setups in the following analysis based on the aforementioned information, and they are shown in in Tab.~\ref{tab_set}. Different exposures of 100 t$\cdot$yr and 1000 t$\cdot$yr are labeled as I and II and different specific setups are labeled with A to D. Set A projects a nominal scenario with a 5\% systematic uncertainty and a similar background level to present xenon direct detection experiments like Xenon-nT and PandaX-4T. Sets B, C and D illustrate optimistic scenarios with 0.5\% systematic uncertainties and different background models. In Set C and D we employed improved background models to project the progression in background discrimination.

\begin{table}
    \centering
    \begin{tabular}{cccc}
        \toprule[1pt]
        Experiment&Exposure&Systematic&Background\\
            Set&[t$\cdot$yr]& Uncertainty& Model\\
        \midrule [1pt]
        I-A&100&5\%&$B_0$\\
        I-B&100&0.5\%&$B_0$\\
        I-C&100&0.5\%&$B_1$\\
        I-D&100&0.5\%&$B_2$\\
        II-A&1000&5\%&$B_0$\\
        II-B&1000&0.5\%&$B_0$\\
        II-C&1000&0.5\%&$B_1$\\
        II-D&1000&0.5\%&$B_2$\\
        \bottomrule[1pt]
    \end{tabular}
    \caption{Simplified experimental scenarios employed in this work. We notate experimental scenarios with exposures of 100 t$\cdot$yr and 1000 t$\cdot$yr with number I and II and the experimental scenarios with different combinations systematic uncertainties and background models are notated with A to D.}
    \label{tab_set}
\end{table}

\section{Numerical Results}
In this section, we present the results of the numerical analysis. First, we present the predicted event spectra for CE$\nu$NS and E$\nu$ES within the Standard Model and in the presence of light mediators, as functions of recoil energy. Next, we provide the independent constraints from different experimental setups on the solar neutrino fluxes and the light mediator models. Finally, we will consider the impact of the light mediators on the measurements of the solar neutrino fluxes across different experimental scenarios.

\subsection{Predicted Event Spectra}
\begin{figure}
    \centering
    \includegraphics[scale=0.6]{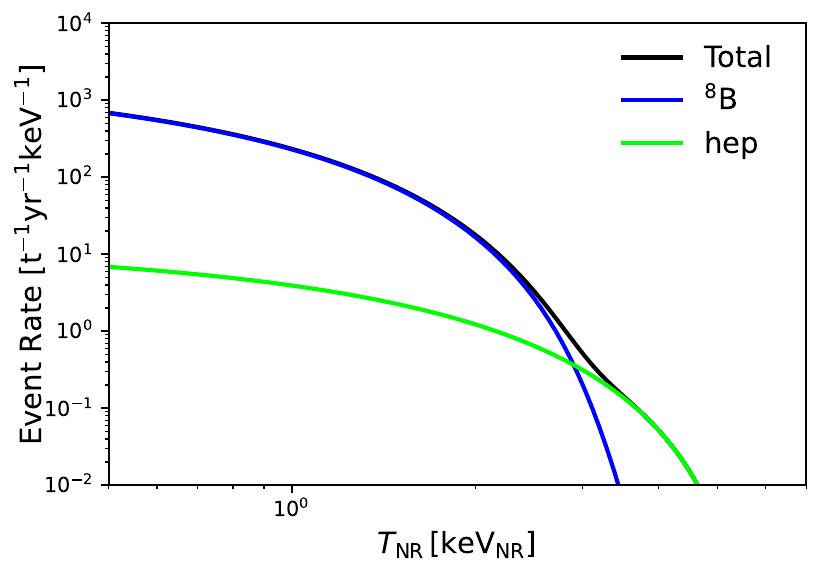}
    \includegraphics[scale=0.6]{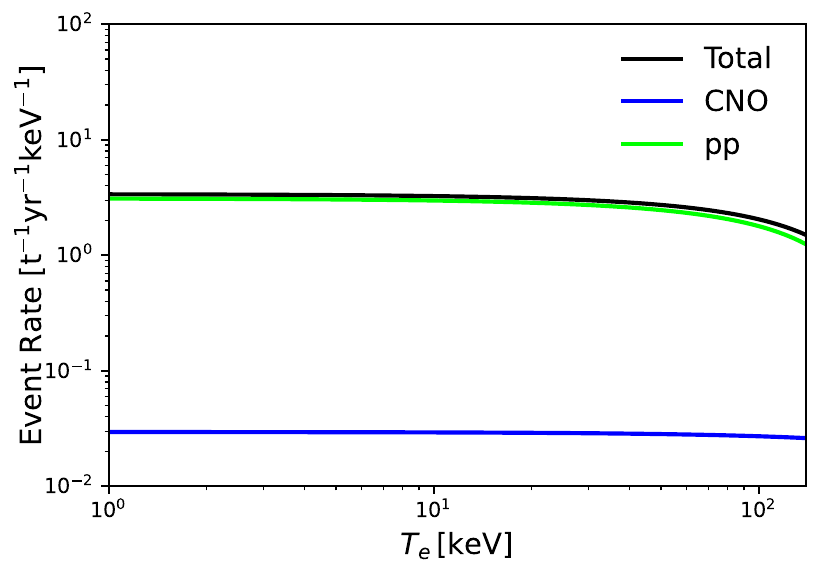}
    \caption{The CE$\nu$NS (left) and E$\nu$ES (right) event rate spectra of solar neutrinos on the liquid xenon target with the standard interaction are presented. In CE$\nu$NS event rate spectrum the two components induced by $^8$B and hep fluxes are shown in blue and green lines respectively. In E$\nu$ES event rate spectrum the dominant component induced by pp flux is shown in green and the total rate of CNO components is also shown in green. A weighted average has been performed based on the natural abundance of isotopes in xenon.}
    \label{spctrum:sm} 
\end{figure}

In Fig.~\ref{spctrum:sm} we illustrate the predicted event energy spectra of the solar neutrino CE$\nu$NS (left) and E$\nu$ES (right) as functions of recoil energy respectively in our ROI. A weighted average has been performed based on the natural abundance of isotopes in xenon. It should be note that only $^8$B and hep neutrinos can induce nuclear recoil in xenon-based direct detection experiments due to their high energy. The $^8$B neutrinos are the dominant contribution to the solar neutrino nuclear recoil events. The hep neutrinos can be detected with nuclear recoil events at about 4 keV$_{\rm NR}$, nearly free from $^8$B neutrino events, enabling the potential measurement of hep fluxes in large-scale direct detection experiments. The pp neutrinos contribute predominantly to electron recoil events in direct detection experiments. However, the uncertainty in the pp neutrino flux could hinder precise measurements on other fluxes, such as CNO flux.

\begin{figure}
    \centering
    \includegraphics[scale=0.6]{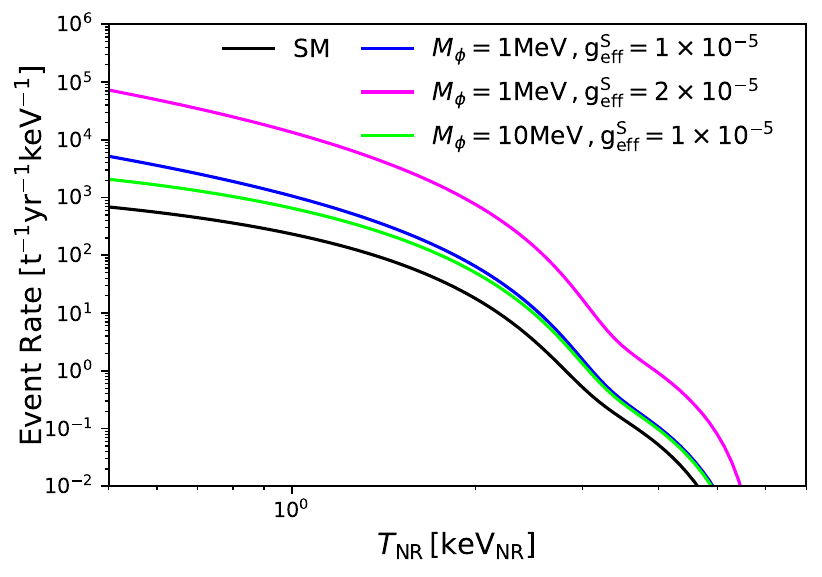}
    \includegraphics[scale=0.6]{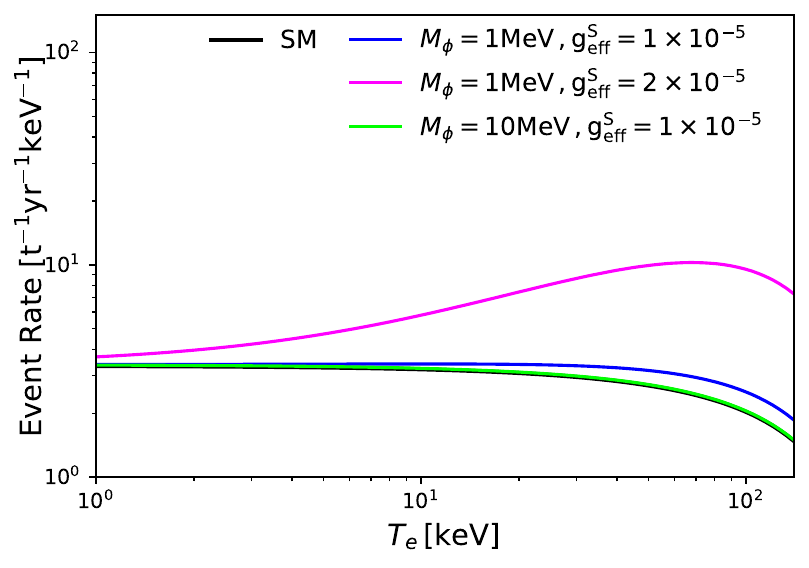}
    \includegraphics[scale=0.6]{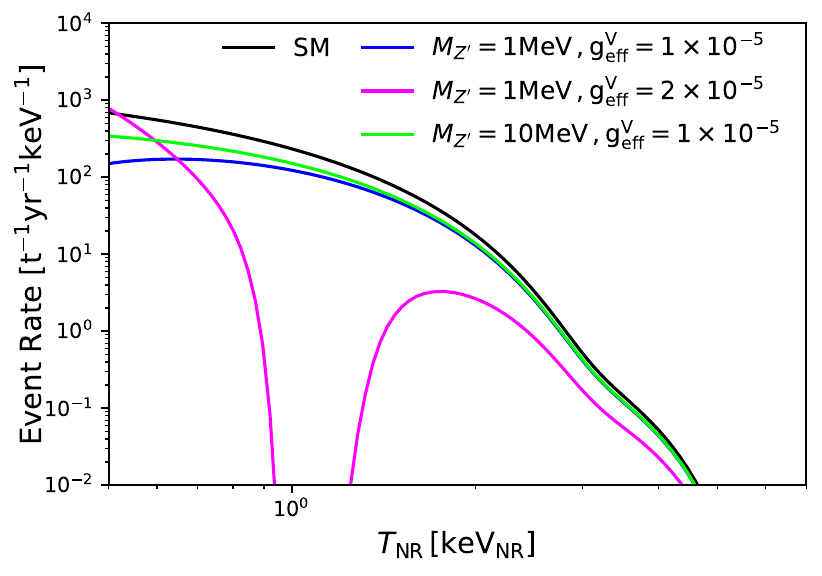}
    \includegraphics[scale=0.6]{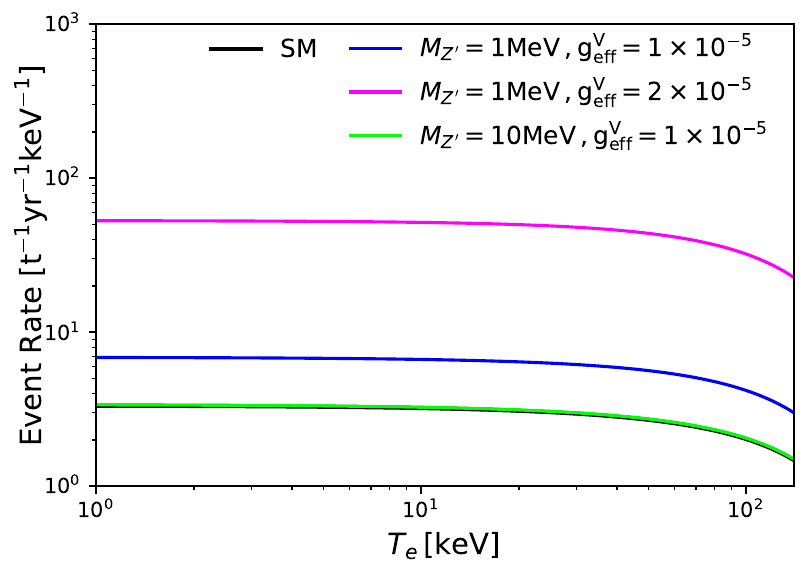}
    \includegraphics[scale=0.6]{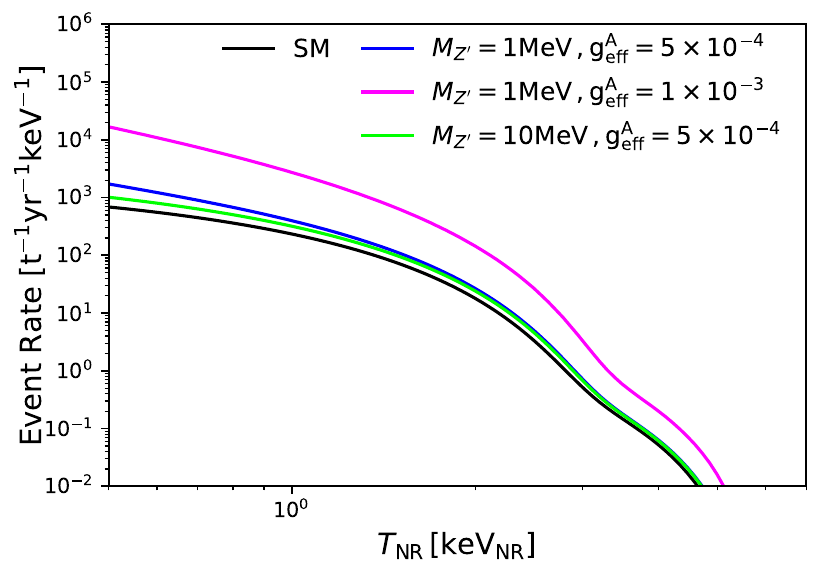}
    \includegraphics[scale=0.6]{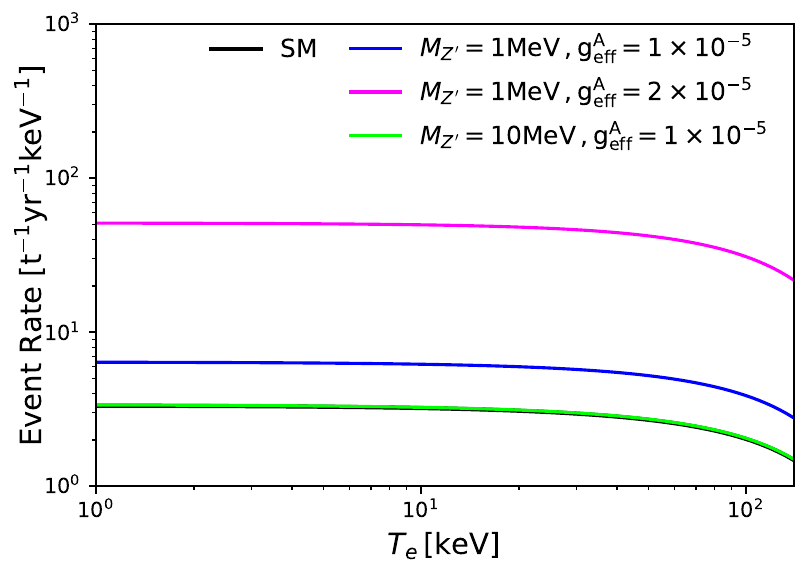}
    \caption{The CE$\nu$NS (left) and E$\nu$ES (right) event rate spectra of solar neutrinos on liquid xenon target in the presence of scalar (top), vector (middle) and axial vector (bottom) mediators are presented. Results based on the Standard Model are shown in black and the results incorporating the non-standard interactions with different parameter sets are shown in different colors. A weighted average has been performed according to the natural abundance of isotopes in xenon.}
    \label{spctrum:nsi} 
\end{figure}

In Fig.\ref{spctrum:nsi} we illustrate the effects of scalar (top), vector (middle) and axial vector (bottom) mediators on the expected event energy spectra of the solar neutrino CE$\nu$NS (left) and E$\nu$ES (right) processes respectively. A weighted average based on isotope abundance is also applied. It should be noted that all the contributions from the light mediators are suppressed by the mass of mediators, as indicated in Eq.~\ref{lagragian:nsi}. By carefully looking into the properties of the figures, several comments on the spectra are provided as follows
\begin{itemize}
    \item For the scalar mediator, since the scalar coupling is incoherent with the standard interaction, the corresponding contribution always enhances the event energy spectra. In the case of CE$\nu$NS, the enhancement dramatically increases as the recoil energy decreases, particularly for a light mediator with a small mass. In the E$\nu$ES process, the scalar mediator enhances the event spectra in a different manner. The cross section continues to increase with the energy in our ROI for a 1 MeV mediator while the total event rate is constrained by the minimum energy required to trigger an electron recoil. Consequently, the event energy spectrum for a 1 MeV mediator and $1\times 10^{-5}$ coupling reaches its maximum value at approximately 100 keV.

    \item In the case of the vector mediator, the contributions of the light mediator will lead to cancellation in the CE$\nu$NS process, resulting in a steep valley on the spectrum for a small mass mediator with a large coupling. However, the contribution of the vector mediator to the E$\nu$ES process is not sensitivity to recoil energy, as the modification to the vector charge remains nearly constant within this energy range for a given set of parameters.

    \item For the axial vector mediator, a large coupling is needed to significantly enhance the CE$\nu$NS event rate since it couples to the spin operator, which is much smaller than the nucleon number in xenon. Similar to the vector mediator, the contribution of the axial vector mediator to E$\nu$ES is also not sensitive to the recoil energy.
\end{itemize}


\subsection{Constraints on the Solar neutrino Fluxes}
\begin{figure}
	\centering
	\includegraphics[scale=0.48]{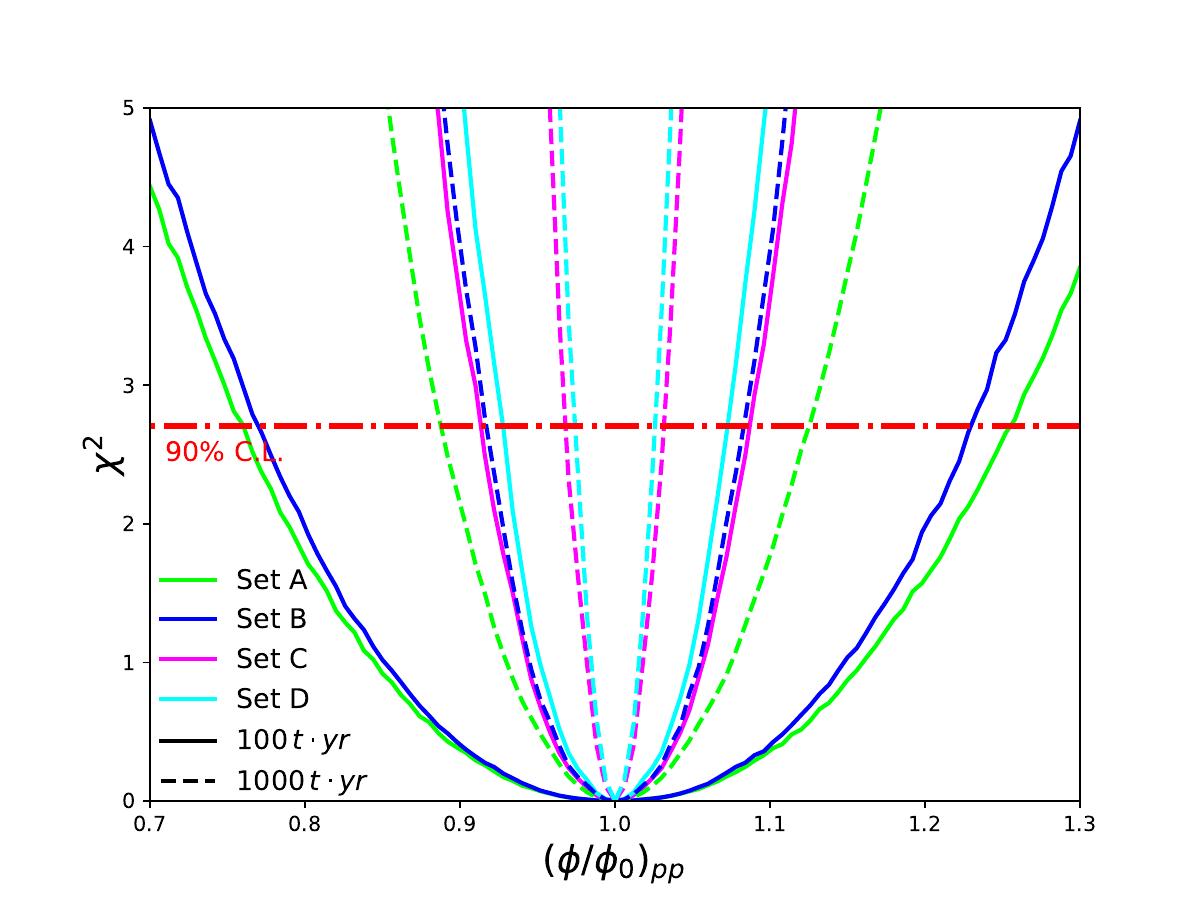}
	\includegraphics[scale=0.48]{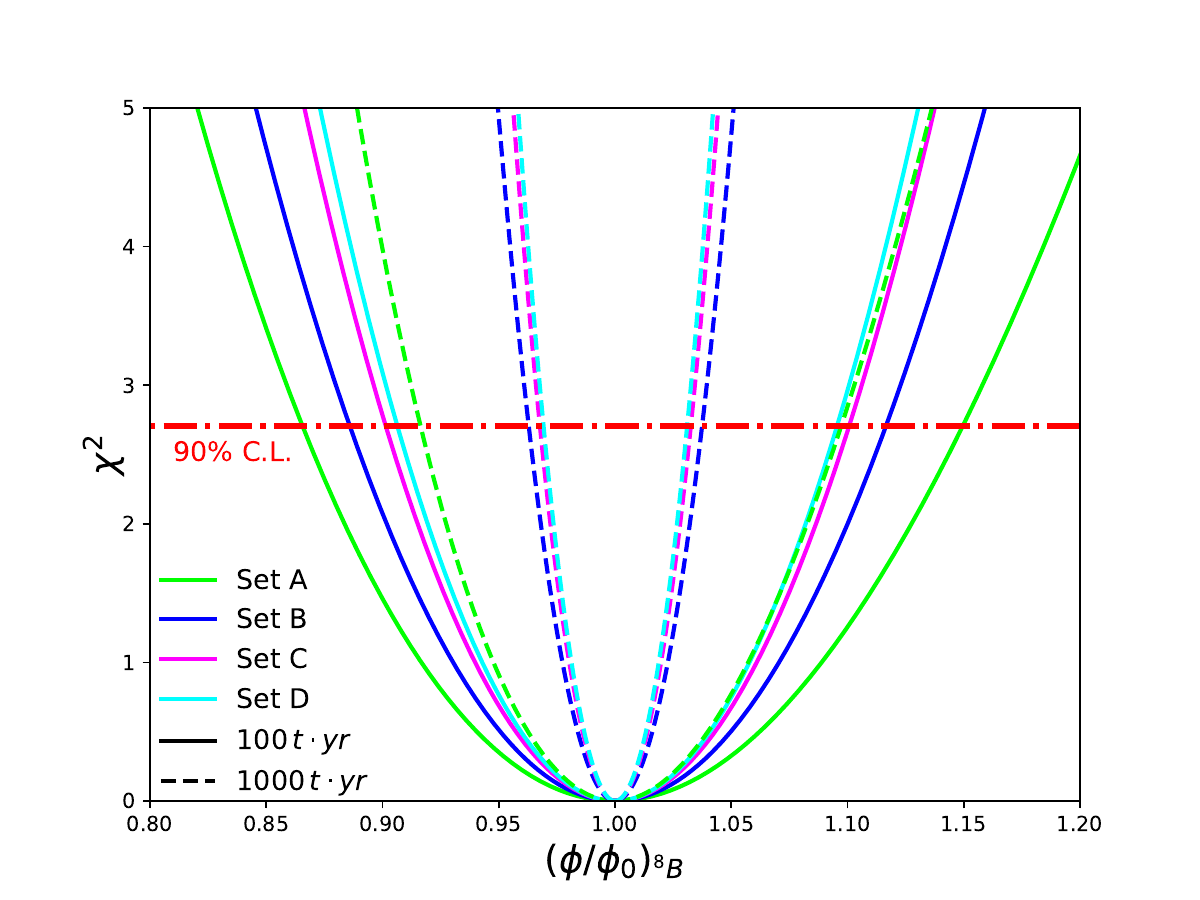}
	\includegraphics[scale=0.48]{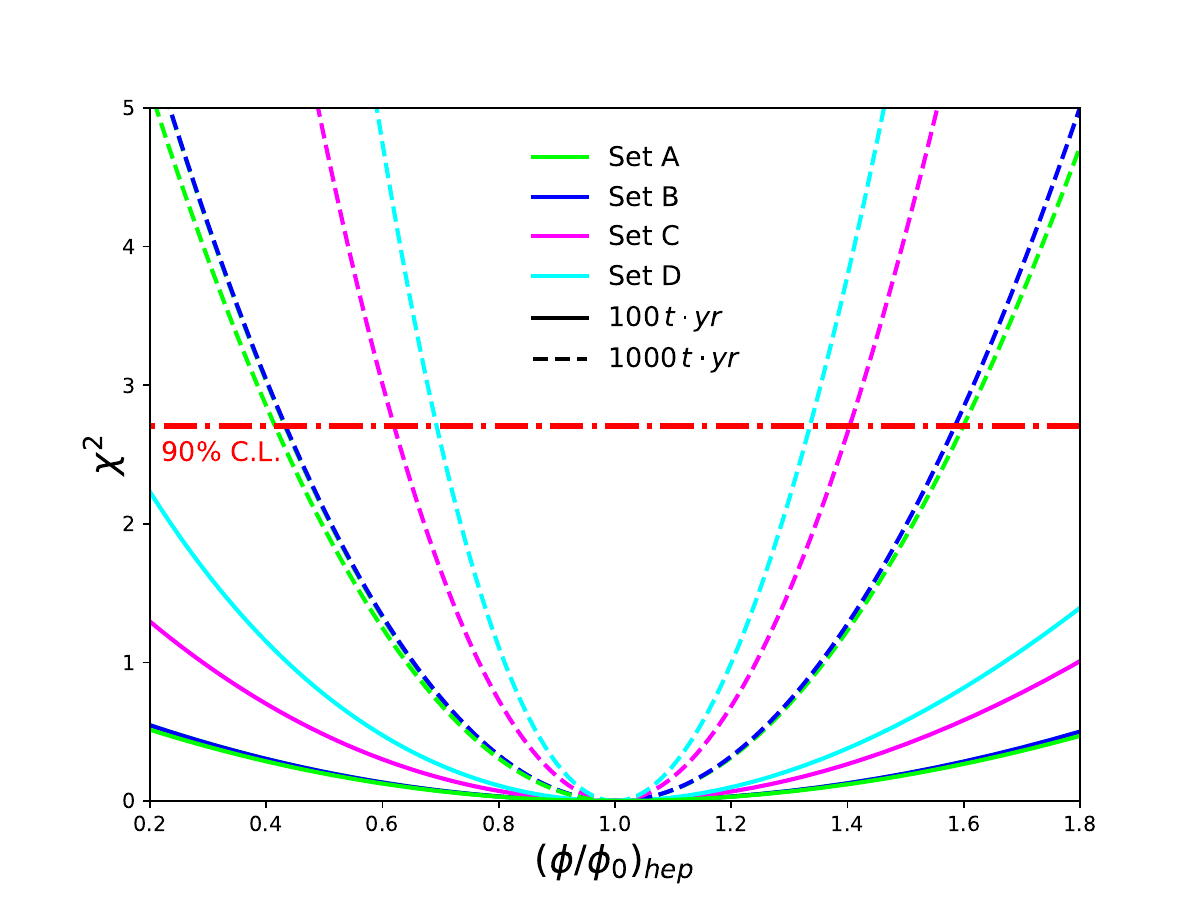}
	\caption{The $\chi^2$ profiles of the pp, $^8$B and hep fluxes with the standard interactions are presented based on the DD experiment sets shown in Tab.~\ref{tab_set} from top to bottom. Profiles corresponding to different uncertainty-background sets are distinguished by different colors. Solid and dashed lines represent profiles for different exposures respectively. The 90\% C.L. cut is indicated by a red dashed line.}
	\label{chi2:flux_sm} 
\end{figure}

In Tab.~\ref{tab_unc_sm} we present the constraints on pp, $^8$B and hep neutrino fluxes at 90\%  confident level (C.L.) in the Standard Model based on the experiment setups shown in Tab.~\ref{tab_set} and the corresponding $\chi^2$ profiles are shown in Fig.~\ref{chi2:flux_sm} from top to bottom. Since $^7$Be, pep and CNO neutrino fluxes can only induce electron recoil in our ROI and pp flux contributes to the dominant electron recoil event numbers, the experiment setups in Tab.~\ref{tab_set} cannot provide effective constraints on $^7$Be, pep and CNO neutrino fluxes due to the uncertainties in the pp flux. The setups with 100 t$\cdot$yr exposure cannot provide effective constraints on hep flux, as shown in Fig.~\ref{chi2:flux_sm}, due to the limited number of events.

It is shown in Tab.~\ref{tab_unc_sm} that xenon based direct detection experiments have great potential in precise measurements of the solar neutrino fluxes. For pp and $^8$B neutrino fluxes, a 100 t$\cdot$yr detector with improved systematic uncertainty and background can achieve 90\% C.L. constraints better than $\pm 10\%$ and in the best experimental setups constraints as good as $\pm 3\%$ can be achieved. For hep neutrino flux, a 1000 t$\cdot$yr exposure detector can achieve constraints better than $\pm 50\%$, provided that background levels are improved.

It should also be noted that the efforts to reduce background can significantly improve the constraints on pp and hep neutrino fluxes, as shown in Fig.~\ref{chi2:flux_sm}, since the corresponding event rates are comparable to or smaller than those of the background model $B_0$. The background model $B_0$ reflects the background levels in current xenon-based direct detection experiments, and thus reducing background in these experiments would be beneficial for advancing measurements of pp and hep neutrino fluxes. For the $^8$B neutrino flux, the corresponding event rate is significantly higher than background rate and a reducing systematic uncertainty will be more effective in improving the constraints, especially when the exposure is sufficiently  high  to detect a large number of events.
\begin{table}
\centering
\begin{tabular}{c|cccccc}
    \toprule[1pt]
    \multirow{2}*{Setup}&\multicolumn{2}{c}{pp}&\multicolumn{2}{c}{$^8$B}&\multicolumn{2}{c}{hep}\\
    ~&$(\Phi/\Phi_0)_{pp}$&$[10^{10}{\rm cm}^{-2}{\rm t}^{-1}]$&$(\Phi/\Phi_0)_{^8 \rm{B}}$&$[10^{6}{\rm cm}^{-2}{\rm t}^{-1}]$&$(\Phi/\Phi_0)_{hep}$&$[10^{4}{\rm cm}^{-2}{\rm t}^{-1}]$\\
    \midrule [1pt]
    I-A&$1^{+0.255}_{-0.240}$&$5.941^{+1.515}_{-1.426}$&$1^{+0.150}_{-0.135}$&$5.20^{+0.78}_{-0.70}$&-&-\\
    I-B&$1^{+0.231}_{-0.230}$&$5.941^{+1.372}_{-1.366}$&$1^{+0.117}_{-0.115}$&$5.20^{+0.61}_{-0.60}$&-&-\\
    I-C&$1^{+0.088}_{-0.088}$&$5.941^{+0.523}_{-0.523}$&$1^{+0.101}_{-0.099}$&$5.20^{+0.53}_{-0.51}$&-&-\\
    I-D&$1^{+0.074}_{-0.073}$&$5.941^{+0.440}_{-0.434}$&$1^{+0.096}_{-0.094}$&$5.20^{+0.50}_{-0.49}$&-&-\\
    II-A&$1^{+0.126}_{-0.113}$&$5.941^{+0.749}_{-0.671}$&$1^{+0.098}_{-0.084}$&$5.20^{+0.51}_{-0.44}$&$1^{+0.60}_{-0.57}$&$3.0^{+1.8}_{-1.7}$\\
    II-B&$1^{+0.083}_{-0.083}$&$5.941^{+0.493}_{-0.493}$&$1^{+0.038}_{-0.038}$&$5.20^{+0.20}_{-0.20}$&$1^{+0.59}_{-0.57}$&$3.0^{+1.8}_{-1.7}$\\
    II-C&$1^{+0.032}_{-0.032}$&$5.941^{+0.190}_{-0.190}$&$1^{+0.033}_{-0.033}$&$5.20^{+0.17}_{-0.17}$&$1^{+0.41}_{-0.39}$&$3.0^{+1.2}_{-1.2}$\\
    II-D&$1^{+0.027}_{-0.026}$&$5.941^{+0.160}_{-0.154}$&$1^{+0.032}_{-0.031}$&$5.20^{+0.17}_{-0.16}$&$1^{+0.34}_{-0.31}$&$3.0^{+1.0}_{-0.9}$\\
    \bottomrule[1pt]
\end{tabular}
\caption{Constraints on pp, $^8$B and hep neutrino fluxes at 90\% C.L. in the Standard Model are presented based on the experiment setups shown in Tab.~\ref{tab_set}. These experiment setups are unable to provide effective constraints on $^7$Be, pep and CNO fluxes. The experiment sets I with a exposure of 100 t$\cdot$yr cannot provide effective constraints on the hep flux due to the very limited event numbers.}
\label{tab_unc_sm}
\end{table}

\subsection{Constraints on the Light Mediators}
\begin{figure}
    \centering
    \includegraphics[scale=0.5]{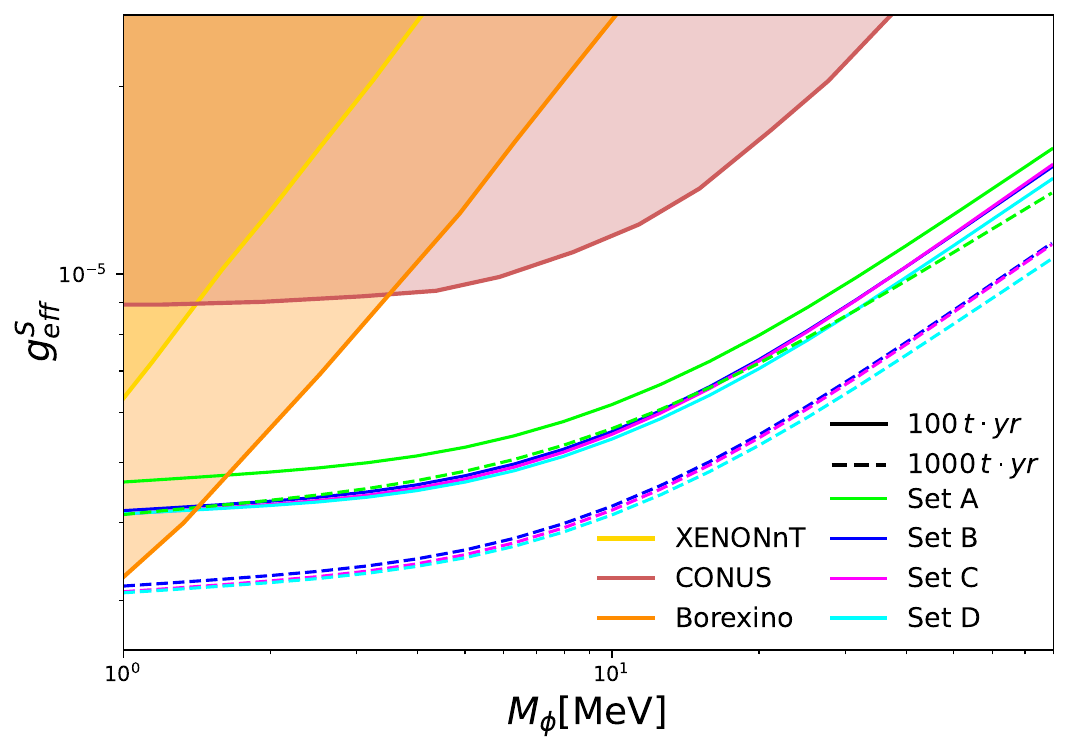}
    \includegraphics[scale=0.5]{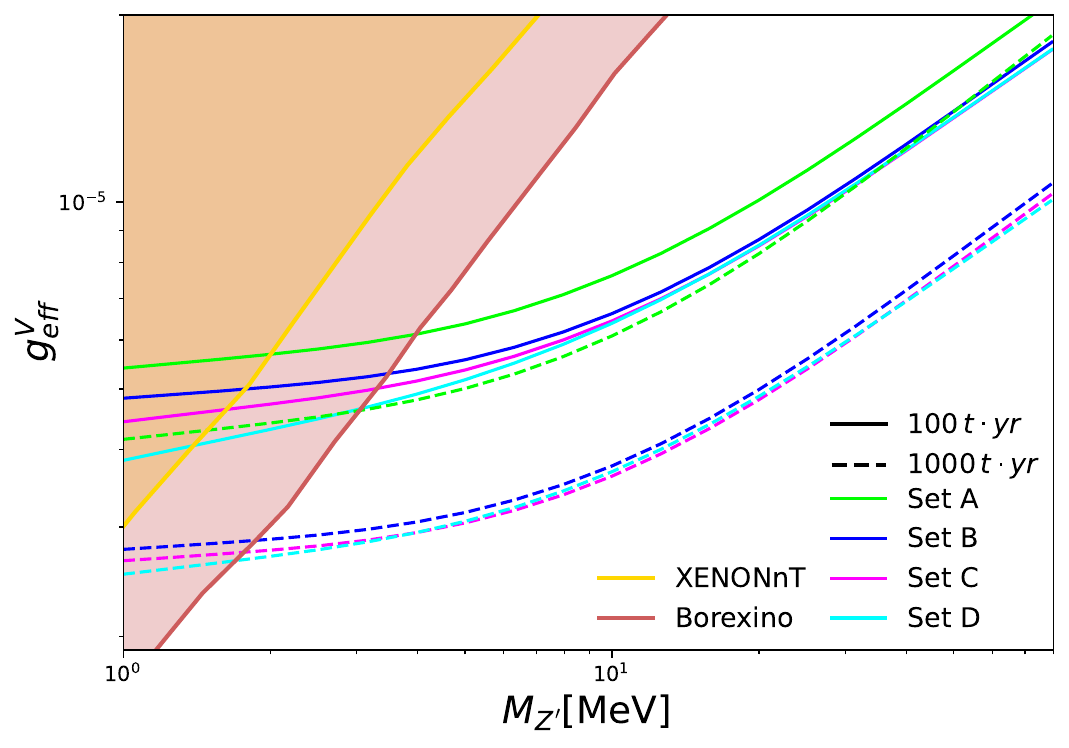}
    \includegraphics[scale=0.5]{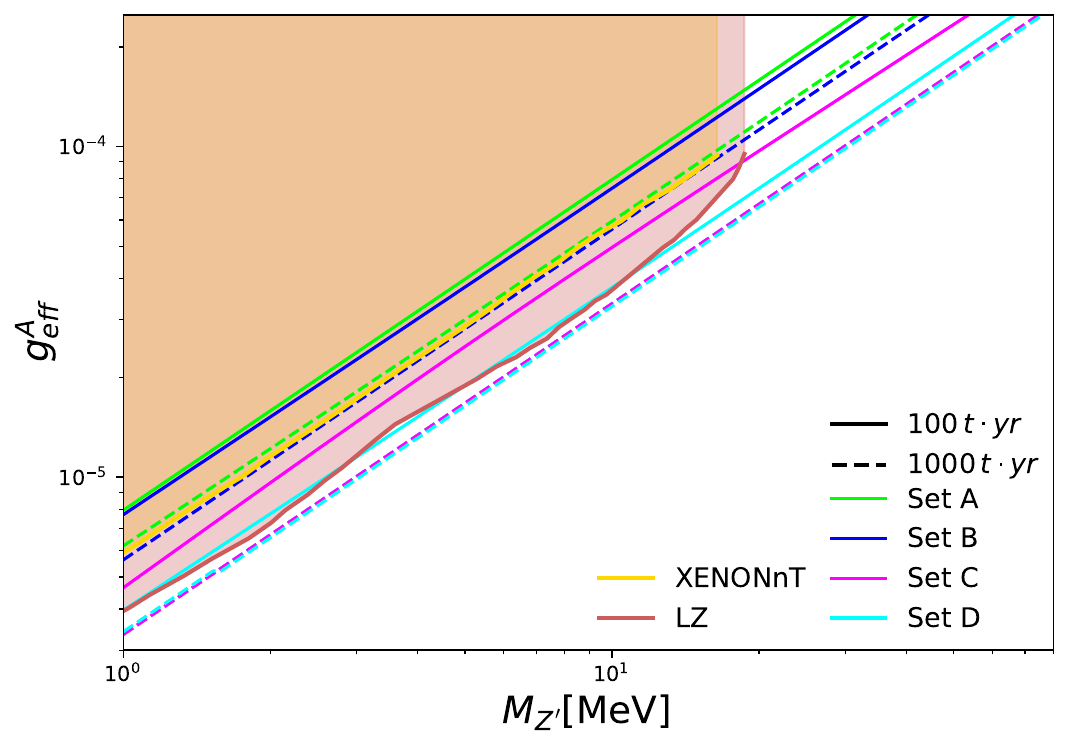}
    \caption{The 90\% C.L. constraints on scalar (top), vector (middle) and axial vector (bottom) mediators are presented based on the constraints on solar neutrino fluxes provided in Tab.~\ref{tab_unc_sm}. Results corresponding to different uncertainty-background sets are distinguished by different colors with solid and dashed lines representing different exposures. We also show the results from XENONnT~\cite{A:2022acy}, LZ~\cite{A:2022acy}, CONUS~\cite{CONUS:2021dwh} and Borexino~\cite{Coloma:2022umy} with shaded regions indicating their respective constraints for comparison.}
    \label{constraint:nsi} 
\end{figure}

In Fig.~\ref{constraint:nsi} we have illustrated the 90\% C.L. upper limits on the parameter space of the scalar (top), vector (middle) and axial vector (bottom) mediators, based on the experimental setups listed from top to bottom in Tab.~\ref{tab_set}. In our analysis, we adopt the constraints from Tab.~\ref{tab_unc_sm} and allow the fluxes without effective constraints to vary freely in order to obtain the independent constraints on light mediators using direct detection experiments. We also present results from XENONnT~\cite{A:2022acy}, LZ~\cite{A:2022acy}, CONUS~\cite{CONUS:2021dwh}, with shaded regions indicating their respective constraints for comparison. In ref.~\cite{A:2022acy}, the authors cite a 7\% uncertainty on the {total} flux~\cite{Serenelli:2016dgz} in their analysis and present promising results with only $\mathcal{O}(1)$ t$\cdot$yr exposure. {However, in this work we consider the independent constraints on light mediators with the direct detection experiments and the uncertainties on the measurements of solar neutrino fluxes shown in Tab.~\ref{tab_unc_sm} weaken the constraints. This highlights} the potential of the combining measurements of the solar neutrinos on both direct detection experiments and other experiments employing different technologies, {which can significantly improve the constraints on the light mediators with better measurements of solar neutrino fluxes.}. Below, we summarize key remarks for each model of the light mediators:
\begin{itemize}
    \item For the scalar mediator, constraints by xenon based direct detection experiments can reach $g^S_{eff}\sim\mathcal{O}(10^{-6})$ when the mediator mass is below 10 MeV. However, the upper limit of the scalar coupling increases rapidly after the mediator mass exceeds 10 MeV, as the contributions from the scalar mediator are suppressed by the mass, as shown in Eq.~\ref{lagragian:nsi}. Improvements in systematic uncertainties lead to significantly more stringent constraints, whereas reductions in background levels have little effect across all mass ranges. The independent constraints from these experiment setups are already better than those from XENONnT, CONUS and Borexino for most mediator mass and they could be further improved with the measurements of solar neutrino fluxes from other experiments employing different technologies and ROIs.
    \item For the vector mediator, the case is similar to that of the scalar mediator. However, improvements in the background levels can push the constraints to a lower level at low mediator masses. At mediator masses as low as $M_{Z^\prime} \lesssim1$ MeV, the cancellation from vector mediator will become so strong that it creates a valley in the CE$\nu$NS event energy spectra at the recoil energy of $\mathcal{O}(1)$ keV$_{\rm NR}$, which is an important feature of the contribution of the vector mediator and needs a low background to observe, especially when the vector coupling is small. As the mediator mass increases, the valley on the event energy spectra moves below the threshold and the vector contribution will also be suppressed, making the reductions in background levels less effective.
    \item For axial vector mediator, the constraints from xenon based direct detection experiments are not as stringent as those for those of scalar and vector mediators and the effects of the improvements in background levels and systematic uncertainties are very limited. As illustrated in Eq.~\ref{lagragian:nsi}, the axial vector mediator coherently couples to spin operator of only $\mathcal{O}(1)$ in CE$\nu$NS, making the corresponding contribution to nuclear recoil event rate much smaller. However, the CE$\nu$NS signals from $^8$B neutrinos contribute significantly to the total event count, while the background in the detection of nuclear recoil is substantially lower than in electron recoil detection. This makes direct detection experiments less effective in constraining the axial vector mediator. 
\end{itemize}

\subsection{Impact of the Light Mediators on the Measurements of the Solar Neutrino Fluxes}
\begin{figure}
    \centering
    \includegraphics[scale=0.45]{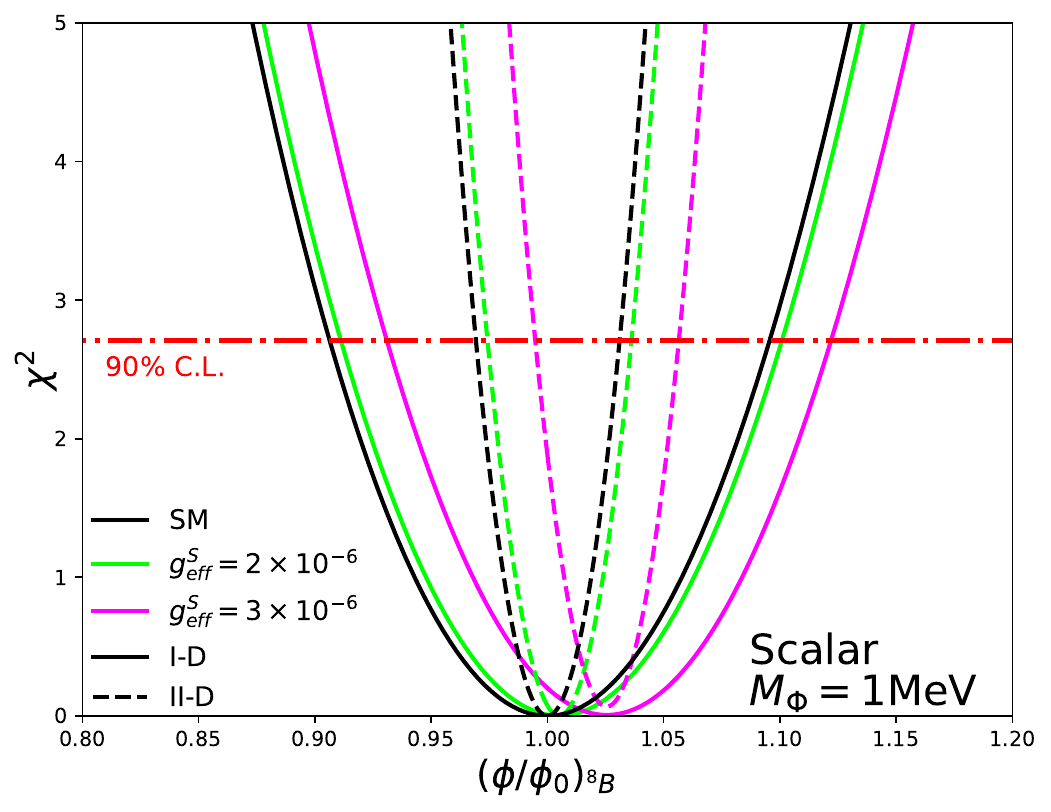}
    \includegraphics[scale=0.45]{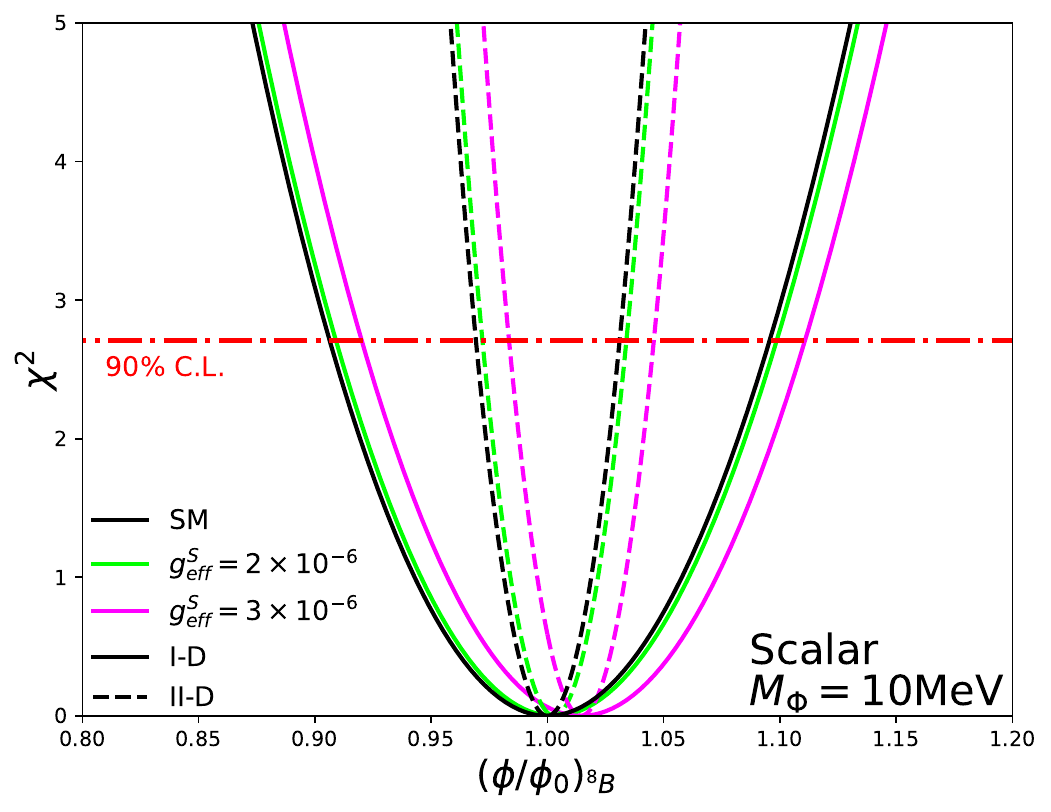}
    \includegraphics[scale=0.45]{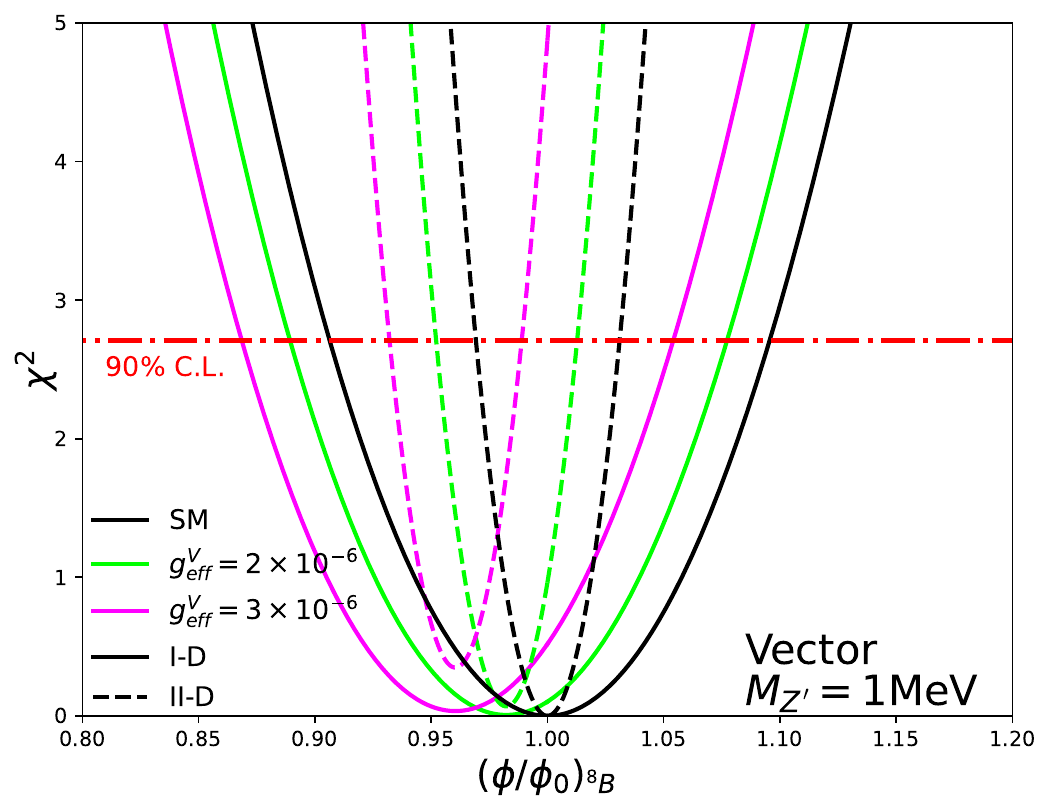}
    \includegraphics[scale=0.45]{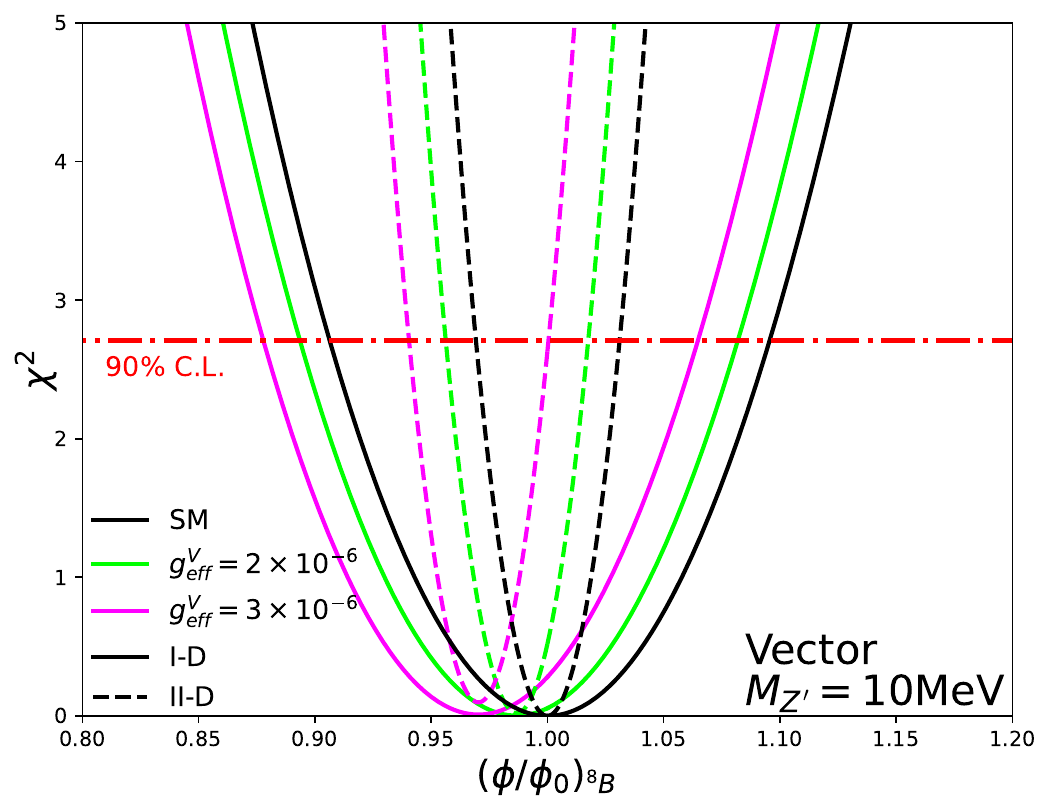}
    \caption{The $\chi^2$ profiles of the $^8$B flux are presented in the Standard Model and in the presence of scalar (top) and vector (bottom) mediators, with mediator masses of 1 MeV (left) and 10 MeV (right). The profiles corresponding to the Standard Model are presented in black line, while those in the presence of light mediators with effective couplings of $2\times10^{-6}$ and $3\times10^{-6}$ are distinguished with green and magenta lines. Solid and dashed lines indicate the profiles based on the experimental setups of I-D and II-D from Tab.~\ref{tab_set} respectively.}
    \label{constraint:nsi_flux} 
\end{figure}

In Tab.~\ref{tab_fit_nsi} we present the best fit values for $^8$B neutrino fluxes in the presence of scalar and vector mediators with varying masses and couplings. The corresponding $\chi^2$ profiles are shown in Fig.~\ref{constraint:nsi_flux}. We consider the experiment setups of I-D and II-D from Tab.~\ref{tab_set}, which represent the most sensitive setups with 100 t$\cdot$yr and 1000 t$\cdot$yr exposures respectively. It was found that $^8$B flux is significantly more sensitive to the light mediators compared to pp and hep fluxes. The light mediators capable of inducing observable effects on pp and hep flux measurements are likely to be excluded by the $\chi^2$ fit to $^8$B fluxes. Additionally, the axial vector components are sub-dominant to the CE$\nu$NS due to the weak couplings and their corresponding nuclear recoil spectra differ significantly from the standard CE$\nu$NS dominated by vector interactions. As a result, the axial vector mediators are likely be excluded by CE$\nu$NS before their couplings become sufficiently strong to induce observable changes in solar neutrino flux measurements. Therefore, this section focuses exclusively on the impact of scalar and vector mediators on the measurement of $^8$B neutrino flux.

Since the presence of the scalar and vector mediators leads to enhancement and cancellation in the $^8$B nuclear recoil events respectively, the best fit values of the $^8$B flux increases with the scalar coupling and decreases with the vector coupling as shown in Tab.~\ref{tab_fit_nsi} and Fig.~\ref{constraint:nsi_flux}. We find that the best fit values are strongly dependent on the coupling strengths, while the mediator masses reduce the shifts in the best-fit values due to the suppression of the corresponding weak charges.Furthermore, we observe that the shifts in the best-fit values cannot be directly mitigated by increased statistics, as shown in Tab.~\ref{tab_fit_nsi}. However, as shown in Fig.~\ref{constraint:nsi_flux}, with increased statistics of 1000 t$\cdot$yr the minimum $\chi^2$ is no longer close to zero for a mediator mass of 1 MeV. Thus, the differences between the contributions of the light mediators and the standard interactions to the spectra are expected to be distinguishable, allowing light mediators to be excluded in future measurements when the mediator mass is small. For the mediators with larger masses, such shifts in the best fit values of $^8$B flux are difficult to mitigate using only independent measurements from direct detection experiments and constraints from other experiments will be necessary to precisely determine the $^8$B flux. Additionally, such features may add to the difficulties in probing light mediators with solar neutrino direct detection experiments, as contributions from light mediators may be interpreted as modifications to the solar neutrino fluxes.

\begin{table}
{
\centering
\begin{tabular}{c|cccccc}
    \toprule[1pt]
    \multirow{2}*{Mediator}&\multicolumn{2}{c}{Scenario}&\multicolumn{2}{c}{I-D}&\multicolumn{2}{c}{II-D}\\
    ~&$M_{\Phi/Z^\prime}$[MeV]&$g^{S/V}_{eff}$&$(\Phi/\Phi_0)_{^8 \rm{B}}$&$[10^{6}{\rm cm}^{-2}{\rm t}^{-1}]$&$(\Phi/\Phi_0)_{^8 \rm{B}}$&$[10^{6}{\rm cm}^{-2}{\rm t}^{-1}]$\\
    \midrule [1pt]
    \multirow{4}*{Scalar}&1&$2\times10^{-6}$&$1.005$&$5.23$&$1.005$&$5.23$\\
    ~&10&$2\times10^{-6}$&$1.003$&$5.22$&$1.003$&$5.22$\\
    ~&1&$3\times10^{-6}$&$1.025$&$5.33$&$1.025$&$5.33$\\
    ~&10&$3\times10^{-6}$&$1.014$&$5.27$&$1.014$&$5.27$\\
    \midrule [1pt]
    \multirow{4}*{Vector}&1&$2\times10^{-6}$&$0.982$&$5.11$&$0.982$&$5.11$\\
    ~&10&$2\times10^{-6}$&$0.987$&$5.13$&$0.987$&$5.13$\\
    ~&1&$3\times10^{-6}$&$0.961$&$5.00$&$0.960$&$4.99$\\
    ~&10&$3\times10^{-6}$&$0.970$&$5.04$&$0.970$&$5.04$\\
    \bottomrule[1pt]
\end{tabular}}
\caption{Best fit values in the presence of the scalar/vector mediators with different masses and couplings under experiment setups of I-D and II-D in Tab.~\ref{tab_set}.}
\label{tab_fit_nsi}
\end{table}
\section{Conclusion}
The dark matter direct detection experiments have entered the multi-ton scale phase and the recent results on the measurements of the solar $^8$B neutrino flux at XENONnT~\cite{XENON:2024ijk} and PandaX-4T~\cite{PandaX:2024muv} show the potential of direct detection experiments as effective platforms for solar neutrino flux measurements. In this work, we present the potential of xenon based direct detection experiments to independently provide precise measurements on the $^8$B, pp and hep fluxes. We also demonstrate the independent sensitivity of xenon based direct detection experiments on the light mediators of scalar, vector and axial vector couplings with CE$\nu$NS and E$\nu$ES events induced by the solar neutrino fluxes. Moreover, we discuss the impact of scalar and vector mediators in the measurements of $^8$B neutrino fluxes. We find that a 100 t$\cdot$yr detector with improved systematic uncertainties and background levels can achieve 90\% C.L. constraints better than $\pm 10\%$ on pp and $^8$B neutrino fluxes and the constraints can be pushed to $\pm 3\%$ with increased statistics. We present that the xenon based direct detection experiments are capable of independently achieving the constraints of $\mathcal{O}(10^{-6})$ on the vector and scalar couplings with the corresponding mediator mass less than 10 MeV. It is also found that the scalar and vector mediators could lead to shifts on the best fit values of the $^8$B neutrino flux and such shifts are difficult to distinguish even with increased experiment scales and technologies, especially when the mediator mass is large. Therefore, this feature will increase the challenges in both the measurement of solar neutrino fluxes and the probing of light mediators in direct detection experiments.

\section*{Acknowledgements}
The author is grateful to Yu-Feng Li for helpful discussions. This work is supported in part by National Natural Science Foundation of China under Grant Nos.~12075255, 12075254 and 11835013.
\newpage
\bibliographystyle{h-physrev5}
\bibliography{main}

\end{document}